\documentclass[usenatbib,fleqn]{mn2e}

\usepackage{amsmath}
\usepackage{mathrsfs}
\usepackage{amssymb}
\usepackage{aas_macros}
\usepackage{graphicx}
\usepackage[dvipsnames]{xcolor}
\usepackage{ulem}

\usepackage[breaklinks,colorlinks,citecolor=blue]{hyperref}
\usepackage[all]{hypcap} 

\newif\iffigs
\newif\iffigscl
\newif\iffigstest
\newif\iflabs

\figsfalse

\figscltrue

\figstestfalse

\labsfalse

\newcommand{\eq}[1]{Eq. (\ref{#1})}

\newcommand{\eqc}[3]{Eq. (\ref{#1}), Eq. (\ref{#2}) and Eq. (\ref{#3})}

\newcommand{\figpan}[1]{{\sc {#1}}}

\newcommand{\dd}{\mathrm{d}}
\newcommand{\arcosh}{\mathrm{arcosh}}

\newcommand{\Myr}{\, \mathrm{Myr}}
\newcommand{\Msun}{\, \mathrm{M}_{_\odot}}
\newcommand{\Pc}{\, \mathrm{pc}}
\newcommand{\Sd}{\, \Msun \, \Pc^{-2}}
\newcommand{\Cmiii}{\, \mathrm{cm}^{-3}}

\newcommand{\ACC}{_{_{\rm ACC}}}
\newcommand{\AMB}{_{_{\rm AMB}}}
\newcommand{\CRIT}{_{_{\rm CRIT}}}
\newcommand{\JEANS}{_{_{\rm J}}}
\newcommand{\subO}{_{_{\rm O}}}
\newcommand{\subS}{_{_{\rm S}}}
\newcommand{\EXT}{_{_{\rm EXT}}}
\newcommand{\MAX}{_{_{\rm MAX}}}
\newcommand{\HALF}{_{_{\rm HT}}}
\newcommand{\DRMAX}{_{_{\rm MAX}}}
\newcommand{\DRFAST}{_{_{\rm FAST}}}
\newcommand{\E}{_{_{\rm E78}}}
\newcommand{\EFAST}{_{_{\rm E78.FAST}}}
\newcommand{\EMAX}{_{_{\rm E78.MAX}}}
\newcommand{\V}{_{_{\rm V83}}}

\newcommand{\VMAX}{_{_{\rm V83.MAX}}}
\newcommand{\W}{_{_{\rm W10}}}

\newcommand{\WMAX}{_{_{\rm W10.MAX}}}

\newcommand{\ATO}{A_{T1}}
\newcommand{\ATT}{A_{T2}}
\newcommand{\BBOT}{_{_{\rm BOT}}}
\newcommand{\CELL}{_{_{\rm CELL}}}
\newcommand{\CMAX}[1]{c_{_{\rm{MAX.}#1}}}
\newcommand{\CMEAN}[1]{\langle c{_{#1}} \rangle}
\newcommand{\EFOLD}{_{_{\rm EFOLD}}}
\newcommand{\EEFOLD}{_{_{\rm E78}}^{_{\rm EFOLD}}}
\newcommand{\EEFOLDFAST}{_{_{\rm E78.FAST}}^{_{\rm EFOLD}}}
\newcommand{\FRG}{_{_{\rm FRG}}}
\newcommand{\HI}{H {\sc i} }
\newcommand{\HII}{H {\sc ii} }
\newcommand{\MCL}{\bar{M}_{\rm CL}}
\newcommand{\NL}{_{_{\rm NL}}}
\newcommand{\NUM}{_{_{\rm NUM}}}
\newcommand{\RAM}{_{_{\rm RAM}}}
\newcommand{\RAD}[1]{_{_{\rm RAD.#1}}}
\newcommand{\SIGTO}{\Sigma_{T1}}
\newcommand{\SIGTT}{\Sigma_{T2}}
\newcommand{\SINK}{_{_{\rm SINK}}}
\newcommand{\ST}{_{_{\rm ST}}}
\newcommand{\VCL}{v_{_{\rm CL}}}

\newcommand{\itii}[1]{{#1}}

\newcommand{\itiitext}[1]{{#1}}

\title[Fragmentation of stratified gaseous layers]{Fragmentation of vertically stratified gaseous layers: monolithic or coalescence--driven collapse}
\author[F. Dinnbier et al.]{Franti\v sek Dinnbier$^{1,2}$\thanks{E-mail:frantisek.dinnbier@asu.cas.cz}, 
Richard W\"unsch$^{1}$, Anthony P. Whitworth$^{3}$ and Jan Palou\v s$^{1}$ \\
$^{1}$Astronomical Institute, Academy of Sciences of the Czech Republic, Bo\v{c}n\'{i} II 1401, 141 00 Prague, Czech Republic \\
$^{2}$Charles University in Prague, Faculty of Mathematics and Physics, V Hole\v{s}ovi\v{c}k\'{a}ch 2, 180 00 Prague, Czech Republic \\
$^{3}$School of Physics and Astronomy, Cardiff University, Cardiff CF24 3AA, Wales, UK}

\begin{document}

\date{Accepted 2016 December 21. Received 2016 December 20; in original form 2016 August 09}

\pagerange{\pageref{firstpage}--\pageref{lastpage}} \pubyear{2017}

\maketitle

\label{firstpage}

\begin{abstract}
We investigate, using 3D hydrodynamic simulations, the fragmentation of pressure-confined, vertically stratified, self-gravitating gaseous layers. 
The confining pressure is either thermal pressure acting on both surfaces, or thermal pressure acting on one surface and ram-pressure on the other. 
In the linear regime of fragmentation, the dispersion relation we obtain agrees well with that derived by \citet{ee78}, 
and consequently deviates from the dispersion relations based on the thin shell approximation \citep{v83} or pressure assisted gravitational instability \citep{wd10}. 
In the non-linear regime, the relative importance of the confining pressure to the self-gravity is a crucial parameter controlling the qualitative course of fragmentation.
When confinement of the layer is dominated by external pressure, self-gravitating condensations are delivered by a two-stage process: 
first the layer fragments into gravitationally bound but stable clumps, and then these clumps coalesce until they assemble enough mass to collapse. 
\itiitext{In contrast}, when external pressure makes a small contribution to confinement of the layer, the layer fragments monolithically into gravitationally 
unstable clumps and there is no coalescence. 
This dichotomy persists whether the external pressure is thermal or ram. 
We apply these results to fragments forming in a shell swept up by an expanding \HII region, and find that, 
unless the swept up gas is quite hot or the surrounding medium \itiitext{has} low density,
the fragments have low-mass ($\la 3\Msun$), and therefore they are unlikely to spawn stars that are 
sufficiently massive to promote sequential self-propagating star formation.
\end{abstract}


\section{Introduction}\label{SEC:INTRO}%

Massive stars ($M_\star\ga 8\Msun$) strongly influence the interstellar medium (ISM) surrounding them, mainly via photoionisation, stellar winds and supernova explosions. 
\citet{el77} propose a mechanism ({\it Collect and Collapse}) whereby an over-pressured \HII region, driven by young massive stars, expands into dense molecular gas. 
The expansion induces a spherical shock, and the surrounding gas accumulates between the shock and the ionisation front. 
The resulting shell of cool gas increases in mass, and eventually fragments to form a new generation of stars. 
A crucial issue is the maximum mass of these newly formed stars. 
If some of them are sufficiently massive to excite a new \HII region, the process can repeat recursively and star formation propagates itself sequentially. 
Otherwise, star formation is quenched. 
Various triggering mechanisms are discussed in \citet{e98}.

From an observational perspective, shells are common morphological structures in the ISM. \citet{ep05} have detected more than 600 shells in \HI. \citet{cw07} have identified 322 complete or partial rings in the infrared. \citet{sp12} list more than 5000 infrared shells. \citet{ds10} find that at least 86\% of the infrared shells identified by \citet{cw07} encircle \HII regions ionised by O- and early B-type stars, suggesting that the shells are due to feedback from these stars. After a careful examination, \citet{dz05} find 17 infrared shells that are candidates for the {\it Collect and Collapse} mechanism. Evidence for propagating star formation has also been reported in significantly larger \HI shells \citep{dm08, dm11, el14}, indicating that feedback operates over a large range of scales. Further observational support for propagating star formation comes from age sequences of OB associations and star clusters in the vicinity of star forming regions e.g. \citep{b64, b91, bg94, bp10} 

In order to estimate the properties of fragments condensing out of a swept up shell, \citet[][hereafter E78]{ee78}, \citet{d80} and \citet{lp93} investigate 
the stability of a vertically resolved self--gravitating layer
\footnote{We distinguish a {\it layer}, which ideally is plane-parallel stratified, from a {\it shell}, 
which ideally is part, or the whole, of a hollow spherically symmetric structure. 
Although most of the observations motivating this work concern shells, 
the configurations we simulate in this paper are layers.
Then we apply results derived for layers in the analysis of star formation triggered by expanding 
\HII regions --- so in that analysis we also use the term shell, because the spatial curvature and velocity divergence of the shell can be neglected.}
%
, and derive a semi-analytical formula for the corresponding dispersion relation. 
\citet[][hereafter V83]{v83} derives the dispersion relation for a self--gravitating infinitesimally thin shell, and this has been 
used to estimate the properties of clumps condensing out of fragmenting shells. 
\citet{wb94b} and \citet{e94} analyse the fragmentation of accreting shells, while \citet{wp01} consider non-accreting shells expanding into a vacuum.

These dispersion relations are based on linear perturbation theory, so they are relevant only as long as the perturbing amplitudes are small. \citet{mn87a, mn87b} extend the linearised solution of \citet{ee78} into the non-linear regime, by including second order terms. They find that a layer
breaks into filaments that become increasingly slender with time. In contrast, \citet{f96} proposes that, in the non-linear regime, the fragments form a semi-regular hexagonal pattern on the surface of the layer. 

In order to test the validity of these analytic derivations, \citet{dw09} obtain a dispersion relation, based on numerical simulations of expanding, non-accreting shells, confined by constant external pressure. They find significant differences between their dispersion relation and that of V83 (based on the thin shell approximation). In subsequent work, \citet[][hereafter W10]{wd10} explain the difference by modifying V83 to include the effect of pressure confinement. They call the mechanism underlying their dispersion relation {\it Pressure Assisted Gravitational Instability} \citep[PAGI][]{wd10}. Since \citet{dw09} simulate fragmentation of a whole shell, and the shell gets thinner with increasing external pressure, they are only able to resolve shells which are confined by low to moderate values of the external pressure, and therefore the W10 dispersion relation is unverified in the case of high external pressure. However, it is in the limit of high external pressure that W10 differs substantially from E78. \citet{vk14} investigate self-gravitating layers permeated by magnetic fields and present a non-magnetic control run. Although their results differ from V83, they are in agreement with both E78 and W10, because their layer is confined by a very low external pressure. \citet{ii11a} also derive a numerical dispersion relation that differs significantly from that of V83.

Our main aims are to simulate the fragmentation of gaseous layers, in both the low and high ambient pressure cases, and to compare the 
results with the analytic or semi-analytic estimates derived in previous studies. 
This comparison addresses three issues: (i) dispersion relations in the linear regime of fragmentation; 
(ii) the elapsed time before the layer \itiitext{forms} gravitationally bound fragments, and the resulting fragment masses; 
and (iii) the possibility that mode interaction leads to fragments distributed on a regular periodically repeating pattern. 
We study only a small square patch on the layer, so that we can attain good resolution in directions perpendicular to the layer, 
even when the layer is significantly compressed. 
In addition to layers confined from both sides by thermal pressure, we also simulate layers accreting onto one surface. 
We use these results to investigate the fragmentation of a shell driven by an expanding 
\HII region --- using an analytic solution to account for the expansion of the \HII region --- and contrast our results with those obtained 
analytically by \citet{wb94b}, and by \citet{ii11a} who performed simulations with small perturbing amplitudes.

The paper is organised as follows. 
Section 2 reviews the most important properties of self--gravitating layers and the analytical estimates (E78, V83, W10) of the dispersion relations.
These estimates are used for comparison with the numerical results. 
Section 3 describes applied numerical methods and initial conditions. 
Section 4 describes simulations of layers confined on both sides by thermal pressure (due to a very hot rarefied gas), and seeded with monochromatic perturbations. 
Section 5 describes simulations of layers confined on both sides by thermal pressure, and seeded with polychromatic perturbations. 
Section 6 describes simulations in which we explore whether fragments forming in layers tend to be arranged into regular patterns.
Section 7 describes simulations of layers in which one side is confined by thermal pressure and the other by the ram pressure of a homogeneous plane-parallel inflow. 
Section 8 considers an \HII region expanding into a homogeneous medium, and estimates the time at which the swept up shell fragments, and the properties of the fragments. 
Section 9 discusses the results, and Section 10 summarises our main conclusions. 
Appendix A 
describes our algorithm developed for finding gravitationally bound fragments.

\section{Linearised theory of layer fragmentation}\label{SEC:DISPRELS}

\subsection[]{The unperturbed state}

\label{sunperturbed}


Before introducing the dispersion relations
describing the growth rate of perturbations in the self-gravitating
presure-confined layer, we review the unperturbed configuration.
The model is also used to generate the initial conditions for our simulations.

We assume that the layer is initially in plane-parallel stratified hydrostatic equilibrium, i.e. it is infinite in the $x$ and $y$ directions, 
its normal points in the $z$ direction, and all quantities are functions of $z$ only. 
The layer is isothermal with sound speed $c\subS$, so the density distribution is as derived by \citet{s42} and \citet{gl65}, i.e.
\begin{equation}\label{sech2}
\rho(z)=\frac{\rho\subO}{\cosh^2(z/H\subO)},
\end{equation}
where $\rho\subO$ is the density on the midplane ($z \!=\!0$),
\begin{equation}\label{h0}
H\subO=\frac{c\subS}{\sqrt{2 \pi G \rho\subO}},
\end{equation}
is the vertical scale height, and $G$ is the gravitational constant. 
The surfaces of the layer are at $z\!=\!\pm z\MAX$, and outside this ($|z|\!>\!z\MAX$) there is a hot gas 
with negligible density which exerts an external pressure $P\EXT$.

The layer is fully characterised by $P\EXT$, $c\subS$ and its surface density $\Sigma\subO$. 
A dimensionless parameter constructed from these quantities \citep{ee78}, 
\begin{equation}\label{a}
A=\frac{1}{\sqrt{1+\left(2 P\EXT/\pi G \Sigma\subO^2\right)}}\,,
\end{equation} 
reflects the relative importance of self-gravity and external pressure in holding the layer together.
Parameter $A$ increases monotonically from nearly 0 (external pressure dominated; $P\EXT \gg  G \Sigma\subO^2$) to 1 (self--gravity dominated).

From pressure equilibrium at the surfaces, $P\EXT\!=\!\rho(z\MAX)c\subS^2$, it follows
\begin{equation}\label{ezmax}
z\MAX=H\subO\arcosh^{-1}\!\left( \frac{\rho\subO c\subS^2 }{P\EXT}\right)=H\subO\arcosh^{-1}\!\left( \frac{1}{\sqrt{1-A^2}}\right)\!.
\end{equation}
Half--thickness of a stratified layer is defined by $H\HALF \equiv \Sigma\subO/(2\rho\subO)$.
Equations (\ref{a}), (\ref{sech2}) and (\ref{h0}) then yield
\begin{equation}\label{ehalfth}
H\HALF = \frac{\Sigma\subO c\subS^2}{2P\EXT+ \pi G \Sigma\subO^2} = \frac{c\subS^2 A^2}{\pi G \Sigma\subO}\,,
\end{equation}
and the midplane density, $\rho\subO$, depends on $\Sigma\subO$ as
\begin{equation}\label{erho0}
\rho\subO=\frac{2P\EXT+ \pi G \Sigma\subO^2}{2 c\subS^2} = \frac{\pi G \Sigma\subO^2}{2 c\subS^2 A^2}.
\end{equation}
Pressure dominated layers are of almost uniform density ($z\MAX \simeq H\HALF$), while self--gravity dominated layers 
have a pronounced density maximum on the midplane ($z\MAX \gg H\HALF$).

\subsection[]{Analytical estimates of the dispersion relation}%


In this section, we review and compare the assumptions underlying the dispersion relations derived by E78, V83 and W10. 
To simplify the discussion, we neglect the effects of shell curvature and velocity divergence, 
by setting the shell radius to infinity. 
The results are therefore applicable to a plane--parallel layer, and can also be applied to a shell, as long as the unstable wavelengths are much shorter than the radius of the shell. 

The dispersion relation gives the perturbation growth rate $\bomega$ as a function of wavenumber $k$. 
In planar geometry, and for the dispersion relations in question, $\bomega$ is either purely real or purely imaginary. 
In this paper, we adopt the convention that growing instability is described by the positive real part of $\bomega$, 
which we denote for simplicity $\omega$, and we omit the oscillating imaginary part.
The characteristic timescale of perturbation growth, its e--folding time, is defined as $t\EFOLD = 1/\omega$.

For all the dispersion relations in question, there is a finite range of unstable wavenumbers 
$(0,k\DRMAX)$, where $k\DRMAX$ is the highest unstable wavenumber.
The maximum growth rate $\omega\DRFAST$ is attained for wavenumber $k\DRFAST$ which is always approximately $k\DRMAX/2$. 
The wavelengths corresponding to $k\DRMAX$ and $k\DRFAST$ are denoted $\lambda\DRMAX$ and $\lambda\DRFAST$, respectively. 
When describing a particular analytical estimate of the dispersion relation, the subscript begins with its name, e.g. $k\EFAST$ is the 
wavelength $k\DRFAST$ for E78. 
The abbreviations E78, V83 and W10 refer to the particular dispersion relation, not to the 
paper where they are firstly described.
The unstable branches of the dispersion relations for both self--gravity ($A = 0.99$) and external 
pressure dominated ($A = 0.18$) layers are shown in the right column of Fig. \ref{fdisp_relat}. 

\iffigscl \begin{figure*}
\includegraphics[width=\textwidth]{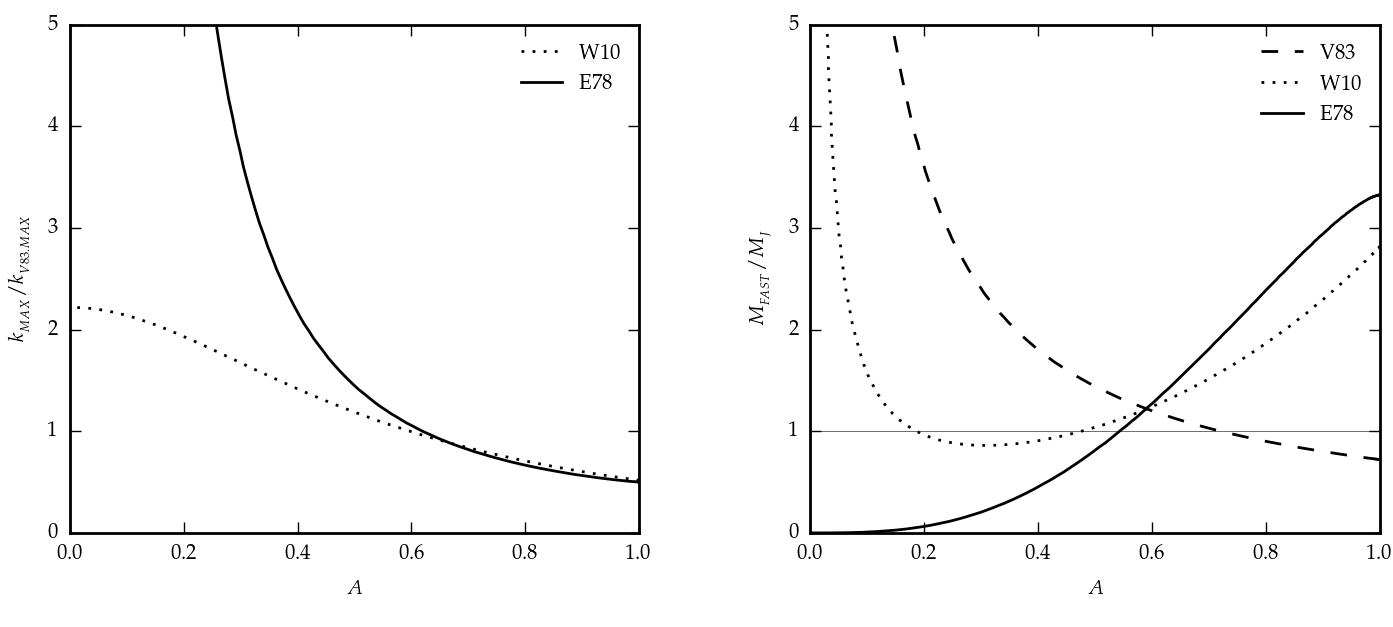}
\caption{Comparison between the dispersion relations derived by E78, V83 and W10, as a function of $A.\;$ \figpan{Left panel:} the marginally stable wavenumber 
$k\DRMAX$, for E78 and W10, normalised to $k\VMAX$.
\figpan{Right panel:} Mass of the fragment with the highest growth rate, $M\DRFAST$, normalised to the midplane Jeans mass, $M\JEANS$. 
Fragments formed with mass below $\sim M\JEANS$ (thin line) are gravitationally stable.}
\label{fdepona}
\end{figure*} \else \fi


\subsubsection{Dispersion relation for the thin shell approximation}%

\itiitext{To estimate the dispersion relation of an expanding shell, \citet{v83} reduce the problem to two dimensions by 
integrating the continuity, Euler, and Poisson's equation through the thickness of the shell.}
In planar geometry, the thin shell dispersion relation becomes 
\begin{equation}\label{tsdr}
\bomega\V^2(k) = 2\pi G \Sigma\subO k - c\subS^2 k^2.
\end{equation}
The stability of modes is determined by the imbalance between self-gravity and internal pressure gradient, and there is no contribution from external pressure.

\subsubsection{Dispersion relation for pressure assisted gravitational instability} %

In order to evaluate the influence of external pressure, \citet{wd10} investigate perturbations with the form of an oblate spheroid, embedded in a layer. 
The spheroid is homogeneous and confined by external pressure $P\EXT$. 
Its semi-major axis is $r$, its semi-minor axis is the layer's half-thickness, $H\HALF$, and its total mass is $M$. 
Radial excursion of an element on the equator of the spheroid is regulated by the equation of motion
\begin{eqnarray}\nonumber
\ddot{r} & = & -\frac{3GM}{2r^2} \left\{ \frac{\cos^{-1}(H\HALF/r)}{(1-(H\HALF/r)^2)^{3/2}} -\frac{H\HALF/r}{1-(H\HALF/r)^2} \right\} \\\label{rdot}
&& -\frac{20 \pi r H\HALF P\EXT}{3 M} + \frac{5 c\subS^2}{r}
\end{eqnarray}
\citep{bw05}.

\citet{wd10} equate the instability growth rate to the radial contraction rate of the spheroid. Collapse from radius $r\subO$, by a small 
factor $\epsilon$ to radius $(1-\epsilon)r\subO$, in time $t_{\epsilon}$, fulfils $(\epsilon - 1)r\subO = \dot{r}\subO t_{\epsilon} + \frac{1}{2} \ddot{r}\subO t_{\epsilon}^2$. 
Using $\bomega=1/t_{\epsilon}$ and $r = \lambda/2 = \pi / k$, this yields
\begin{eqnarray}
\bomega\W^2(k)\!&\!=\!&\!-\frac{\ddot{r}\subO}{2 \epsilon r\subO} \nonumber \\
\!&\!=\!&\!\frac{1}{\epsilon} \bigg\{ - \frac{5 c\subS^2 k^2}{2 \pi^2} \nonumber \\ 
& & + \frac{3 G \Sigma\subO k}{4} \!\left(\!\frac{\cos^{-1}(kH\HALF/\pi)\!-\!\sqrt{k^2 H\HALF^2 /\pi^2\!-\!1}}{(1-k^2 H\HALF^2 /\pi^2)^{3/2}}\right) \nonumber \\
& & + \frac{10 P\EXT c\subS^2 k^2}{3 \pi^2 (2  P\EXT+\pi G \Sigma\subO^2)}  \bigg\}. \label{pagidr}
\end{eqnarray}
The terms on the right hand side of \eq{pagidr} represent the gradient of internal pressure (second line), self-gravity (third line), 
and confinement by external pressure (fourth line). 
\citet{wd10} suggest setting $\epsilon\!\sim\!0.1$, but, although $\epsilon$ affects the magnitude of $\bomega\W$, it does not influence the range of unstable wave-numbers. 

\subsubsection{Dispersion relation for a vertically stratified layer}%

\citet{ee78} obtain the dispersion relation for a self-gravitating, pressure confined, 
semi-inifinite layer in hydrostatic equilibrium by solving the system of perturbed continuity, Euler and Poisson equations. 
\citet{kk12} revisit their study and offer an additional insight.
According to \citet{kk12}, the dispersion relation can be written 
\begin{equation}\label{tddr}
\bomega\E^2 = 2 \pi G \Sigma\subO k (\eta F_{_{\rm J}}+(1-\eta)F_{_{\rm D}}) - c_{_{\rm EFF}}^2 k^2 \,.
\end{equation}
Here $c_{_{\rm EFF}}$ is the effective sound speed. $F_{_{\rm J}}$ and $F_{_{\rm D}}$ are reduction factors for self-gravity. 
Parameter $\eta$ is the fraction of the perturbed surface density that is attributable to compression of material near the centre of a proto-fragment 
(and hence near the mid-plane of the layer). 
The rest of the perturbed surface density is due to corrugations on the surface of the layer. 
Thus $\eta$ controls the relative importance of compressional and surface-gravity waves. 
\itiitext{The former are important in self--gravity dominated layers, while the latter are important in pressure dominated layers. 
$c_{_{\rm EFF}}$, $\eta$, $F_{_{\rm J}}$ and $F_{_{\rm D}}$ are complicated functions of $k$, which can not generally be expressed in closed form.}

The growth rate $\omega\EFAST$ of the most unstable wavenumber $k\EFAST$ in the limit $A \to 0$ (layers dominated by external pressure) 
is (eq. (47) in \citet{kk12})
\begin{equation}\label{ee78olim}
\omega\EFAST^2 =  0.276 \times 2 \pi G \rho\subO= 0.276 \left( \frac{\pi G \Sigma\subO}{c\subS A} \right)^2.
\end{equation}


\subsubsection{Comparison between the analytical estimates of the dispersion relation}

The left panel of Figure \ref{fdepona} shows the dependence of $k\DRMAX$ normalised to $k\VMAX$ on parameter $A$ for \itiitext{different} dispersion relations. 
For self--gravity dominated layers, both W10 and E78 predict almost identical ranges of unstable wavenumbers ($k\EMAX/k\WMAX \to 0.966$ as $A \to 1$), 
but V83 predicts a broader range with $k\WMAX/k\VMAX \to 0.518$. 
With increasing external pressure (decreasing $A$), the difference between E78 and V83 diminishes.
However, when $A\la 0.5$, E78 extends to much higher wavenumbers, indicating that shorter wavelengths are unstable. 
And, as $A\to 0$, W10 and E78 have different limiting behaviour: $k\WMAX$ tends to a constant ($k\WMAX/k\VMAX \to 2.216$), 
whereas $k\EMAX \propto H\HALF^{-1} \propto A^{-2}$. 

Since the wavelength $\lambda\DRFAST$ has the highest growth rate, 
we assume that the fragment mass $M\DRFAST$ is approximately the mass confined inside a circle of radius $\lambda\DRFAST/2$, 
i.e. $M\DRFAST = \pi \Sigma\subO (\lambda\DRFAST/2)^2 = \pi \Sigma\subO (\pi/k\DRFAST)^2$.
Masses $M\DRFAST$ normalised to the midplane Jeans mass $M\JEANS$ are plotted in the right panel of Fig. \ref{fdepona}. 
V83 and W10 predict that when formed, the fragments are already gravitationally unstable ($M\DRFAST/M\JEANS \ga 1$) for any $A$. 
On the other hand, E78 predicts gravitationally unstable fragments only when $A \ga 0.5$.
According to E78, the layer breaks into gravitationally stable fragments for lower values of $A$, thus predicting qualitatively different scenario than V83 and W10.
We simulate and discuss evolution of pressure dominated layers (low $A$) in Sections \ref{snonlinear} and \ref{sdpagi}.

\section{Methods and initial conditions}\label{SEC:METH}%

\subsection{Numerics}\label{SEC:NUM}%


All the hydrodynamic simulations presented in the paper are performed with the MPI-parallelised {\sc flash4.0} code \citep{fo00}. 
{\sc flash} is an {\sc amr} code based on the {\sc paramesh} library \citep{mo00}. 
The hydrodynamic equations are solved by the piecewise parabolic method \citep{cw84}. 

Self-gravity is calculated using an octal tree code (W\"unsch et al. in preparation) which offers three acceptance criteria for interaction between the 
target point (where the acceleration is evaluated) and a node (the source of the gravitational force):
\begin{enumerate}
\item
The algorithm invented by \citet{bh86}, which accepts nodes seen from the target
point at an angle smaller than a specified value. This algorithm is purely
geometric since it does not take into account distribution of mass inside nodes, 
or their relative contribution to the net gravitational force.
\item
A set of criteria using the node size, mass and optionally higher multipole
moments of the mass distribution within the node. They either estimate the upper
limit on the acceleration error of the cell-node interaction  $\Delta
a^\mathrm{max}$ from Equation~9 of \citet{sw94}, or they use the approximation by \citet{s05},
\begin{equation}
\Delta a_\mathrm{(p)}^\mathrm{max} = \frac{GM}{d^2}\left(\frac{h}{d}\right)^{p+1},
\label{ape_mac}
\end{equation}
where $M$ and $h$ are the node mass and size, respectively, $d$ is the distance
between the cell and the node mass centre, and $p$ is the order of the multipole
expansion.
\item
An experimental implementation of the "sumsquare" criterion of
\citet{sw94}, which controls the sum of all errors origination from all individual node contributions.
\end{enumerate}
We invoke the criterion (ii) with Equation (\ref{ape_mac}) and $p = 1$ in all
our simulations because our tests show that it provides the best compromise
between the code performance and accuracy.

We convert dense gaseous condensations into sink particles according to conditions described in \citet{fb10}. 
To create a sink particle inside a particular cell, all following conditions must be fulfilled:
\begin{enumerate}
\item
the cell is at the highest refinement level,
\item
the cell contains minimum of gravitational potential,
\item
all the gas inside a sphere of accretion radius $r\ACC$ centered at the cell is above a specified density threshold,
\item
the gas inside the sphere is gravitationally bound, Jeans-unstable and converging,
\item
radius $r\ACC$ of the sink particle would not overlap with accretion radius of an already existing sink particle.
\end{enumerate}
Following \citep{fb10}, we set the accretion radius to 2.5 grid cell size at the highest refinement level.

All the simulations have mixed boundary conditions (BCs) for self-gravity, i.e. periodic in the two directions 
$x,y$ parallel to the layer, and isolated in the third direction $z$ perpendicular to the layer. 
In order to calculate the gravitational field in this configuration, it is natural to seek for a modification of the standard Ewald method \citep{e21, k97}.
By computing the appropriate limit of the standard Ewald method, we find formulae for the gravitational acceleration and potential in closed 
form for a configuration with mixed BCs; the derivation is described in (W\"unsch et al. in preparation).
The modification is used to calculate the gravitational field in the simulations presented here.

The hydrodynamic BCs are periodic in the $x$ and $y$ directions, and reflecting in the $z$ direction for all but accreting simulations. 
For accreting simulations, the BCs are inflow from the top of the computational domain, and diode from the bottom to prevent reflections of waves.

The column density needed for cooling the warm ambient gas intermixed with the cold layer during accreting runs 
(see Section \ref{smab} for details) is calculated using module \texttt{TreeRay/OpticalDepth} of the tree code.

The grid cells are cubic in all \itiitext{the} simulations.
The half-thickness of the layer ($H\HALF$; see Eqn. \ref{ehalfth} below), is always at least four grid cells (see Tables \ref{tmono} to \ref{tan}). 
Consequently, the Jeans length is always resolved by more than four grid cells, and the simulations 
satisfy the criterion for avoiding artificial gravitational fragmentation, as given by \citet{tk97}. 
We have performed successful convergence tests throughout the range of conditions simulated. 
We have also checked that the portion of the layer inside the computational domain is sufficiently large, 
i.e. that the periodic copies in the $x$ and $y$ directions do not significantly influence the properties of fragments.

\subsection{Initial conditions for the layer}\label{SEC:ICs}%

Each model starts with a layer with properties described in Section
\ref{sunperturbed} seeded with a perturbation. The perturbation is either a
single mode given by the eigenfunction for acoustic-surface-gravity modes 
or white noise. The exact form of the
perturbation is described at the beginning of a corresponding Section
(\ref{smono} to \ref{saccreting}).

Particular values for the layer parameters are adopted from \citet{ii11a} who 
investigate an \HII region excited by a $41 \Msun$ star,
expanding into a medium with number density $n \!=\! 10^3\,{\rm cm}^{-3}$.
At time $t = 0.81\,{\rm Myr}$, the shell has radius $R = 3.86\,{\rm pc}$, and surface
density $\Sigma\subO = 0.0068\,{\rm g}\,{\rm cm}^{-2}$.
We use this value of $\Sigma\subO$ for the initial conditions of all the simulations presented here, and vary $A$ by changing $P\EXT$.
However, we note that the model is sufficiently simple that its physical parameters,
$(\Sigma\subO, c\subS, P\EXT)$, can be rescaled arbitrarily, as long as $A$ is unchanged, i.e. $P\EXT \propto \Sigma\subO^2$.

The sound velocity, $c\subS$, is related to the temperature $T$ by the ideal gas law
$c\subS^2\!=\!\gamma R_{gas} T / \mu$, where $\gamma$ is the effective barotropic exponent, $R_{gas}$ is the ideal gas constant,
and $\mu$ is the mean molecular weight.
Inside the layer, we set $\gamma = 1.0001$ (effectively isothermal, $\gamma = 1.0$ is excluded with the adopted numerical scheme)
$\mu = 2$ (pure molecular hydrogen), and $T = 10\,{\rm K}$.

\subsection{The external medium}\label{smab}%


We use two different kinds of the layer confinement: in Sections
\ref{smono}, \ref{spoly} and \ref{smodint}, we investigate layers confined with
thermal pressure from both surfaces; in Section \ref{saccreting}, we investigate
layers confined with the ram pressure from one surface, and with thermal
pressure from the other.

The medium imposing the thermal pressure is implemented as follows.
In order to compare our simulations with the analytic theory, it is necessary that the ambient medium has no dynamical influence on the layer, apart from exerting the thermal pressure. 
By means of convergence tests we establish that this can be achieved by setting
$\rho\AMB(z\MAX) \lesssim 10^{-2} \rho\subO$. 
In addition, to diminish the influence of sound waves reflection from borders of the computational domain, we extend the computational domain to $\sim \pm 3 z\MAX$.
Therefore, we set the ambient medium to be isothermal with temperature
$T\AMB=300\,{\rm K}$ and mean molecular weight $\mu\AMB = 0.6$.
To prevent the ambient medium from falling on the layer, we set its density
profile close to the hydrostatic equilibrium, i.e. 
\begin{eqnarray}
\rho\AMB(z) & = & \rho\AMB(z\MAX) \times \nonumber \\
 & & \exp \left(-\frac{2 \pi G \Sigma_0 \mu\AMB (|z| - z\MAX)}{R_{gas} T\AMB} \right).
\end{eqnarray}
%


In the case of accreting layers, ram pressure is realised by supersonic accretion of gas with uniform density $\rho\ACC$, temperature $T\ACC$ and sound velocity $c\ACC$.
The gas impacts at velocity $v\ACC$ (and hence Mach number $\mathscr{M} = v\ACC/c\subS$) onto the upper surface of the layer at $z \simeq z\MAX$.
The accreted gas is cold and molecular and the shock is isothermal, i.e. $T\ACC=T$ and $c\ACC=c\subS$.
The bottom surface at $z \simeq - z\MAX$ is confined by the ambient medium at temperature $T\AMB$, $T\AMB \gg T$ exerting the thermal pressure on the layer. 
To prevent the layer from bulk acceleration, we set the pressures acting on both surfaces to be equal, i.e. $\rho\ACC (v\ACC^2 + c\ACC^2) = \rho\AMB c\AMB^2$.

The accretion leads to large scale flows inside the layer (see Section \ref{saccretinglin} and Fig. \ref{fslice_m08}), 
which mix the layer with the warmer ambient medium. 
Consequently, if cooling were not implemented, the temperature of the layer would increase and the layer would thicken. 
This is a spurious behaviour which we suppress because a real layer would quickly cool to temperature $T$. 

To keep the layer at constant temperature, we need to distinguish it from the warm ambient medium. 
We detect the layer according to the surface density $\sigma$ calculated from the bottom side $z\BBOT$ of the computational domain
\begin{equation}
\sigma(z)=\int\limits_{z\BBOT}^{z}\,\rho(z')\,dz'\,.
\end{equation}
We cool to temperature $T$ any cell with the column density above threshold $\sigma\CRIT$.
We set $\sigma\CRIT = 2 \int\limits_{z\BBOT}^{-z\MAX}\,\rho\AMB(z')\,dz'$ to enable the layer to freely ripple.
The layer detection is sensitive because the total column density of the ambient medium is of the order of the column density of one cell inside the layer, 
so $\sigma(z)$ rises steeply once $z$ enters the layer. 

Although most of the mass delivered to the layer comes from the accreted medium, 
the intermixing consumes a significant amount of the warm ambient medium. 
To prevent the ambient medium from being exhausted in the course of a simulation, we continuously 
replenish gas at temperature $T\AMB$ through the bottom boundary of the computational domain, 
so that the total mass of ambient gas remains constant.
%
%
The fresh ambient gas moves towards the layer and induces two artificial
effects: a small ram pressure acting on the contact discontinuity and an increase of layer surface density $\Sigma$. 
We discuss the influence of the ram pressure at the end of Section \ref{saccnl} and show that it is not important.


%


\section[Layers with monochromatic perturbations]{Layers confined by thermal pressure with monochromatic perturbations}

\label{smono}


\itii{In this section, we test the dispersion relations by comparing them to simulations.}
We study two extreme cases of pressure confinement: self--gravity dominated with $A = 0.99$ and external pressure dominated with $A = 0.18$. 
The applied perturbation is of a single wavelength (monochromatic).

Since the analytical estimates are based on linearised equations, they are valid only as 
long as the perturbing amplitude $q_1$ of any quantity is smaller than its unperturbed value $q\subO$. 
Accordingly, we define the linear regime of fragmentation if the maximum of the perturbed surface density $\Sigma_1$ is 
smaller than $\Sigma\subO$ and the non--linear regime otherwise. 
The dispersion relation can be determined only in the linear part of the fragmenting process.

\iflabs
\textit{Generic names}
\else\fi
The generic name of a monochromatic simulation is in the form M${<}A{>}$$\_$${<}kH\HALF{>}$, where the first two numbers 
after "M" represent the value of parameter $A$ and the numbers after the underscore the perturbing wavenumber in the dimensionless form $kH\HALF$. 
\itiitext{Thus, for example, simulation M18\_020 treats a layer with $A = 0.18$, and initial monochromatic perturbation $kH\HALF = 0.20$.}

\subsection{Initial conditions for perturbations}

\iflabs
\textit{Initial perturbations for monochromatic simulations}
\else\fi
\itiitext{The initial monochromatic perturbation for a layer confined by thermal pressure}
corresponds to the eigenfunction for acoustic--surface--gravity modes of 
a thick layer with freely moving surfaces (see eqs. (20) -- (23) in \citet{kk12}).
The initial amplitude of perturbed surface density is $\Sigma_{1} (0) = 0.01 \Sigma\subO$. 
The length of the computational domain in direction $x$ is equal to one perturbing wavelength.

\subsection{The dispersion relation}

\label{smonodr}

\begin{table*}
\begin{tabular}{ccrcccc|c}
Run & $A$ & $n_x \times n_y \times n_z$ & $H\HALF/dz$ & $kH\HALF$  & $t\EEFOLD$ & $t\NUM$ \\
 & &  &  &  & [Myr] &  [Myr] \\
\hline
M18\_020 &  0.18 &   640 $\times$   128 $\times$   128 &  20.2 & 0.20 & 0.16 & 0.16         \\
M18\_025 &  0.18 &   512 $\times$   128 $\times$   128 &  20.2 & 0.25 & 0.15 & 0.16         \\
M18\_033 &  0.18 &   384 $\times$   128 $\times$   128 &  20.2 & 0.33 & 0.15 & 0.16         \\
M18\_050 &  0.18 &   256 $\times$   128 $\times$   128 &  20.2 & 0.50 & 0.19 & 0.19         \\
M18\_062 &  0.18 &   208 $\times$   128 $\times$   128 &  20.2 & 0.62 & 0.41 & 0.41         \\
M18\_073 &  0.18 &   176 $\times$   128 $\times$   128 &  20.2 & 0.73 &    - &    -         \\
M18\_089 &  0.18 &   144 $\times$   128 $\times$   128 &  20.2 & 0.89 &    - &    -         \\
M99\_025 &  0.99 &   512 $\times$   128 $\times$   128 &  20.2 & 0.25 & 0.74 & 0.76         \\
M99\_033 &  0.99 &   384 $\times$   128 $\times$   128 &  20.2 & 0.33 & 0.69 & 0.70         \\
M99\_050 &  0.99 &   256 $\times$   128 $\times$   128 &  20.2 & 0.50 & 0.67 & 0.68         \\
M99\_073 &  0.99 &   176 $\times$   128 $\times$   128 &  20.2 & 0.73 & 0.77 & 0.78         \\
M99\_089 &  0.99 &   144 $\times$   128 $\times$   128 &  20.2 & 0.89 & 1.12 & 1.14         \\
M99\_114 &  0.99 &   112 $\times$   128 $\times$   128 &  20.2 & 1.14 &    - &    -         \\
M99\_133 &  0.99 &    96 $\times$   128 $\times$   128 &  20.2 & 1.33 &    - &    -         \\
\end{tabular}
\caption{\itiitext{Parameters for simulations of layers confined by thermal pressure with monochromatic perturbations (models M).} 
We list parameter $A$, number of cells at the highest refinement
level $n_x$, $n_y$ and $n_z$, resolution in the vertical direction $H\HALF/dz$, parameter $kH\HALF$ where $k$ is the selected 
wavenumber, $t\EEFOLD$ is the analytical e--folding time according to E78 for given $k$, and $t\NUM$ ($t\NUM = 1/\omega$) 
is the e--folding time measured in our simulations.}
\label{tmono}
\end{table*}

\iffigscl
\begin{figure*}
\includegraphics[width=\textwidth]{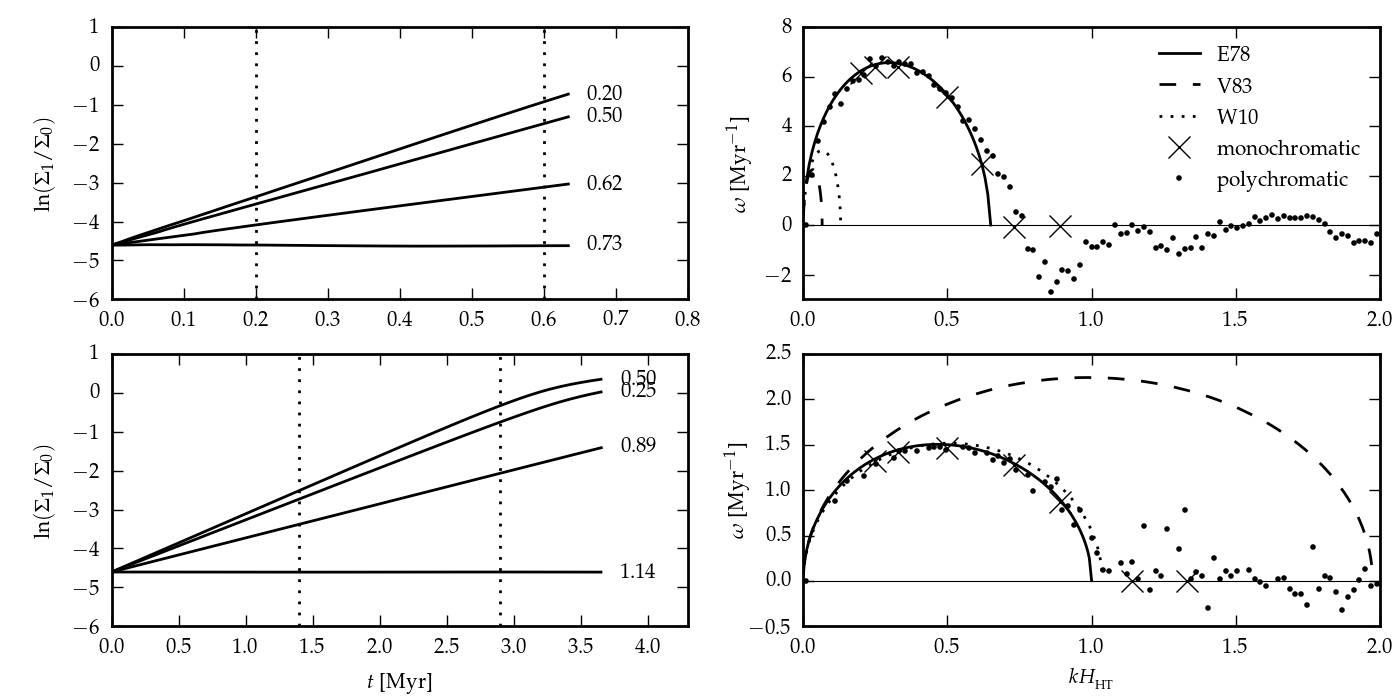}
\caption{The dispersion relation for a self--gravity dominated ($A = 0.99$; top row) and external--pressure dominated ($A = 0.18$; bottom row) layer.
\figpan{Left panels:} Time evolution of the surface density perturbations for monochromatic models.
Only selected models are plotted to avoid confusion. The value of parameter $kH\HALF$ for a particular model is
on right from the curve. The growth rate is calculated in
the interval marked by the vertical dotted lines.
\figpan{Right panels:} Comparison between analytical (E78, V83 and W10) and numerically
obtained dispersion relations.
The growth rate obtained from monochromatic and polychromatic simulations is plotted by crosses and dots, respectively.
Data for polychromatic simulations is binned to reduce noise.
See Sections \ref{smonodr} and \ref{spolydr} for a detailed description.}
\label{fdisp_relat}
\end{figure*}
\else
\fi

\iflabs
\textit{monochromatic sims for pressure dominated case}
\else\fi
Simulations for the external pressure dominated layer ($A = 0.18$, models M18) are listed in Table \ref{tmono}.
Upper left panel of Figure \ref{fdisp_relat} shows evolution of surface density perturbations for selected modes.
Since a small perturbing amplitude in the surface density behaves as $\Sigma_1 (t) = \Sigma_{1} (0) \mathrm{e}^{i \omega t}$, 
the instantaneous growth rate $\omega$ equals to the slope of the curves shown in the upper left panel of Fig. \ref{fdisp_relat}.
$\omega$ is almost time independent throughout the simulations.
Modes with $kH\HALF \la 0.65$ are unstable and grow, modes with $kH\HALF \ga 0.65$ are stable.
We measure $\omega$ by linear fit to $\ln {(\Sigma_1 / \Sigma_{1} (0))}$ over time interval $t^{fit}\subO$, $t^{fit}_1$.
The starting time $t^{fit}\subO$ is determined so as to suppress the influence of the initial conditions (which are exactly that of E78 eigenvectors). 
We suppose that the initial growth rate should be significantly altered at the timescale 
of sound crossing time through one half of the wavelength, i.e. $\simeq \pi H\HALF/c\subS$ (we use $t^{fit}\subO = 0.1 \Myr$).
The upper bound $t^{fit}_1$ is constrained by the condition for the perturbation to be small, i.e. $\Sigma_1 < \Sigma\subO$.
Since model M18 terminates before this condition is fulfilled, we set $t^{fit}_1$ near the end of the simulation.
We compare measured $\omega$ with the analytical dispersion relations \eqc{tsdr}{pagidr}{tddr} in the upper right panel of Figure \ref{fdisp_relat} 
and list measured e--folding time $t\NUM = 1.0/\omega$ alongside the E78 analytical estimate $t\EEFOLD$ in Table \ref{tmono}. 
Our results are very close to E78 (solid line) and are inconsistent with both W10 (dotted line) and V83 (dashed line).

\iflabs
\textit{monochromatic sims for self--gravity dominated case}
\else
\fi
Monochromatic simulations for the self--gravity dominated layer ($A = 0.99$, models M99) also show nearly time independent $\omega$ (lower left panel of Figure \ref{fdisp_relat})
with deviations towards the end only when the perturbing amplitude becomes large $\Sigma_1 \ga \Sigma\subO$.
Measured $\omega$ ($t^{fit}\subO = 1.4 \Myr$, $t^{fit}_1 = 3.0 \Myr$) is in a good agreement with E78 dispersion relation and inconsistent with V83 (lower right panel of Fig. \ref{fdisp_relat}).
Since W10 and E78 are similar, our data could not distinguish between them for self--gravity dominated layers.


\section[Layers with polychromatic perturbations]{Layers confined by thermal pressure with polychromatic perturbations}

\label{spoly}

In this Section, we investigate fragmentation of self--gravitating pressure confined layers both in linear and non--linear regime.
Initial perturbations contain many wavelengths simultaneously (polychromatic).
In the linear regime of fragmentation, we measure the dispersion relation.
We also study the fragmenting process qualitatively as a function of parameter $A$, and compare 
the fragment masses and fragmenting timescales with analytical estimates.

%
%
\iflabs
\textit{Generic names}
\else\fi
The generic name of a polychromatic simulation consists of letter "P" followed by two numbers indicating the value of the parameter $A$. 
\itiitext{For example, simulation P18 treats a layer with $A = 0.18$.}

\subsection{Initial conditions for perturbations}

\label{spolyinit}

\iflabs
\textit{Initial perturbations for white noise simulations}
\else\fi
\itiitext{To initiate simulations of layers confined by thermal pressure with polychromatic perturbations,} 
we perturb the layer with many wavevectors pointing in all three spatial directions.
In order to be able to perform resolution tests, we generate
their amplitudes in the Fourier space and then map them by the inverse Fourier transform
on a grid, so their spectrum does not depend on grid resolution. 
The amplitudes $\tilde{A}(k)$ are drawn as random variables from the uniform distribution
for $\|\vec{k}\| < k\subO$ and are zero otherwise, so the amplitudes occupy a sphere in the Fourier space.
To assure reasonable resolution, the highest wavenumber $k\subO$ corresponds to the size of at least 4 grid cells.
The whole unstable range of wavenumbers predicted by any of the discussed estimate to the dispersion relation 
is always included inside the sphere, i.e. $\mathrm{max} (k\EMAX, k\VMAX, k\WMAX) < k\subO$.

\iflabs
\textit{comparison between models with different A}
\else\fi
In order to compare fragmenting timescales between layers with different values of parameter $A$, it is necessary to
choose comparable amplitudes of their initial perturbation.
To reach the aim, we normalise all the perturbing amplitudes $\tilde{A}(k)$ to satisfy $\sqrt{\langle \tilde{A}(k)^2 \rangle}/\tilde{A}(0) = const$.
\itiitext{To be able to calculate the dispersion relation in the simulations, the initial perturbing amplitudes must be small.
This imposes an upper limit on the normalisation constant.}
The lower limit is constrained by accuracy of the tree code. 
\itiitext{To fulfill these requirements, we set $\Sigma_1 \simeq 0.1 \Sigma\subO$.}


\subsection{The dispersion relation}

\label{spolydr}

\iflabs
\textit{White noise simulations analysis}
\else\fi
To measure the dispersion relation, we compute Fourier transform of surface density for any frame of
the simulation and then fit growth rate of each individual mode in time interval $(t^{fit}\subO$, $t^{fit}_1)$. 
As the initial conditions do not correspond to the eigenfunctions, the modes relax at the beginning, 
and need some time before start growing at a temporarily constant growth rate. 
The choice of $t^{fit}\subO$ is constrained so as the modes with the longest wavelengths 
present in the computational domain already grow at a constant rate. 
$t^{fit}_1$ is constrained so as the modes with the highest growth rate have not reached the nonlinear regime yet.
We find that $t^{fit}\subO = 1/\omega\EFAST$ and $t^{fit}_1 = 3/\omega\EFAST$ fulfill 
at best the requirements for our parameter choice.

\iflabs
\textit{white noise simulations in both regimes}
\else\fi
We calculate the dispersion relation for the same values of parameter $A$ (models P18 and P99 in Table \ref{twn}; right panels of Figure \ref{fdisp_relat}) as 
for the monochromatic models presented in Section \ref{smono}.
The data are binned to reduce noise.
The critical wavenumber $k\MAX$ as well as the growth rate of wavenumbers $k < k\MAX$ are again in
a very good agreement with E78 both in self--gravity and external pressure dominated cases. 
\itiitext{Wavenumbers with $k > k\MAX$ are stable.}
We do not detect any significant deviation between our results and E78.

\subsection{Evolution in non--linear regime}

\label{snonlinear}

\begin{table*}
\begin{tabular}{ccrcccccccccc}
Run & $A$ & $n_x \times n_y \times n_z$ & $H\HALF/dz$ & & $N\JEANS$ & $M\JEANS$ & $\MCL$ & $t\EEFOLDFAST$ & $t\FRG$ & $t\SINK$ & $t\NL$  & \\
 & & & &  &  &  [$\Msun$] & [$\Msun$] & [Myr] & [Myr] &  [Myr] & [Myr] & \\
\hline
P18 &  0.18 &  1024 $\times$  1024 $\times$    64 &   4.0 & &   8.8 &   2.3 &   1.1 &  0.15 &  2.01 &  2.15 &  0.42 \\
P20 &  0.20 &  1024 $\times$  1024 $\times$    64 &   4.4 & &   9.3 &   2.5 &   1.9 &  0.17 &  2.19 &  2.40 &  0.50 \\
P22 &  0.22 &  1024 $\times$  1024 $\times$    64 &   4.4 & &  12.8 &   2.8 &   1.7 &  0.18 &  2.11 &  2.23 &  0.58 \\
P25 &  0.25 &  1024 $\times$  1024 $\times$    64 &   4.4 & &  19.0 &   3.1 &   3.1 &  0.21 &  1.90 &  2.20 &  0.80 \\
P30 &  0.30 &   512 $\times$   512 $\times$    64 &   3.9 & &   9.8 &   3.6 &   2.1 &  0.25 &  2.20 &  2.40 &  0.99 \\
P40 &  0.40 &   512 $\times$   512 $\times$    64 &   4.7 & &  17.0 &   5.0 &   3.5 &  0.33 &  2.38 &  2.64 &  1.45 \\
P60 &  0.60 &   512 $\times$   512 $\times$    64 &   6.6 & &  29.7 &   7.4 &  10.0 &  0.47 &  3.00 &  3.35 &  2.22 \\
P80 &  0.80 &   512 $\times$   512 $\times$    64 &   5.9 & &  89.0 &   9.9 &  20.6 &  0.59 &  3.60 &  4.00 &  2.31 \\
P99 &  0.99 &   512 $\times$   512 $\times$    64 &   9.1 & &  71.0 &  12.5 &  32.0 &  0.66 &  4.15 &  4.90 &  2.90 \\
\end{tabular}
\caption{
\itiitext{Parameters for simulations of layers confined by thermal pressure with polychromatic perturbations (models P).}
First four columns have the same meaning as the columns in Table \ref{tmono}.
Further, we provide number of Jeans masses in the computational domain $N\JEANS$, Jeans mass $M\JEANS$,
mean mass of gravitationally bound objects $\MCL$ at fragmenting time $t\FRG$,
time $t\SINK$,
analytical e--folding time $t\EEFOLDFAST$ for the most unstable wavenumber $k\DRFAST$ and transition time between linear and non--linear regime $t\NL$.}
\label{twn}
\end{table*}

\iflabs
\textit{self--gravity dominated case, monolithic collapse}
\else\fi
Evolution of surface density for self--gravity dominated model P99 is shown in Figure \ref{fsgr_frag}.
The fragmentation begins with emergence of round objects (at time from 0.7Myr to 2.1Myr). These objects then
gradually grow and become slender as was predicted by the second order perturbation theory by \citet{mn87b}.
The transformation of objects from roundish to filamentary--like is apparent between plots at  2.8Myr to 4.9Myr.
Sink particles form in the densest parts of the filaments and accrete material from surrounding filaments. Comparing the last
plot at 4.9Myr with a plot at the early stage of fragmentation (e.g. 2.1Myr), we see that when a clump appears,
it monolithically collapses to a gravitationally unstable object.

\iflabs
\textit{pressure dominated case, two stage collapse}
\else\fi
Fragmentation in pressure--dominated case is represented by model P18 (Fig. \ref{fep_frag}). 
At the beginning of fragmentation, the layer swiftly breaks into small 
objects (plots from 0.3Myr to 1.0Myr) with masses smaller than the Jeans mass $M_J$. 
The subjeans masses are a direct consequence of fragmentation according to E78 for layers with low $A$ (right panel of Fig. \ref{fdepona}).
For confinement of the objects, external pressure is more important than self--gravity, so 
apart from being immersed in an external gravitational field of the layer, the objects are equivalents to gravitationally stable Bonnor--Ebert spheres. 
The stable objects then gradually merge until enough mass for a gravitationally bound clump is assembled (see plots from 1.0Myr to 2.3Myr). 
Merging often leads to non--radial accretion resulting in spinning--up the fragments and disc formation around them (plot at time 2.3Myr).
As a bound clump is formed, it collapses and its cross section for possible following mergers is reduced and merging rate decreases. 
Therefore the fragmenting process in the pressure dominated case, which proceeds via coalescence of many small clumps, is qualitatively 
different from the continuous collapse in the self--gravity dominated case. 
We refer to the former and latter as \textit{coalescence driven collapse} and \textit{monolithic collapse}, respectively.

\iffigscl
\begin{figure*}
\includegraphics[width=\textwidth]{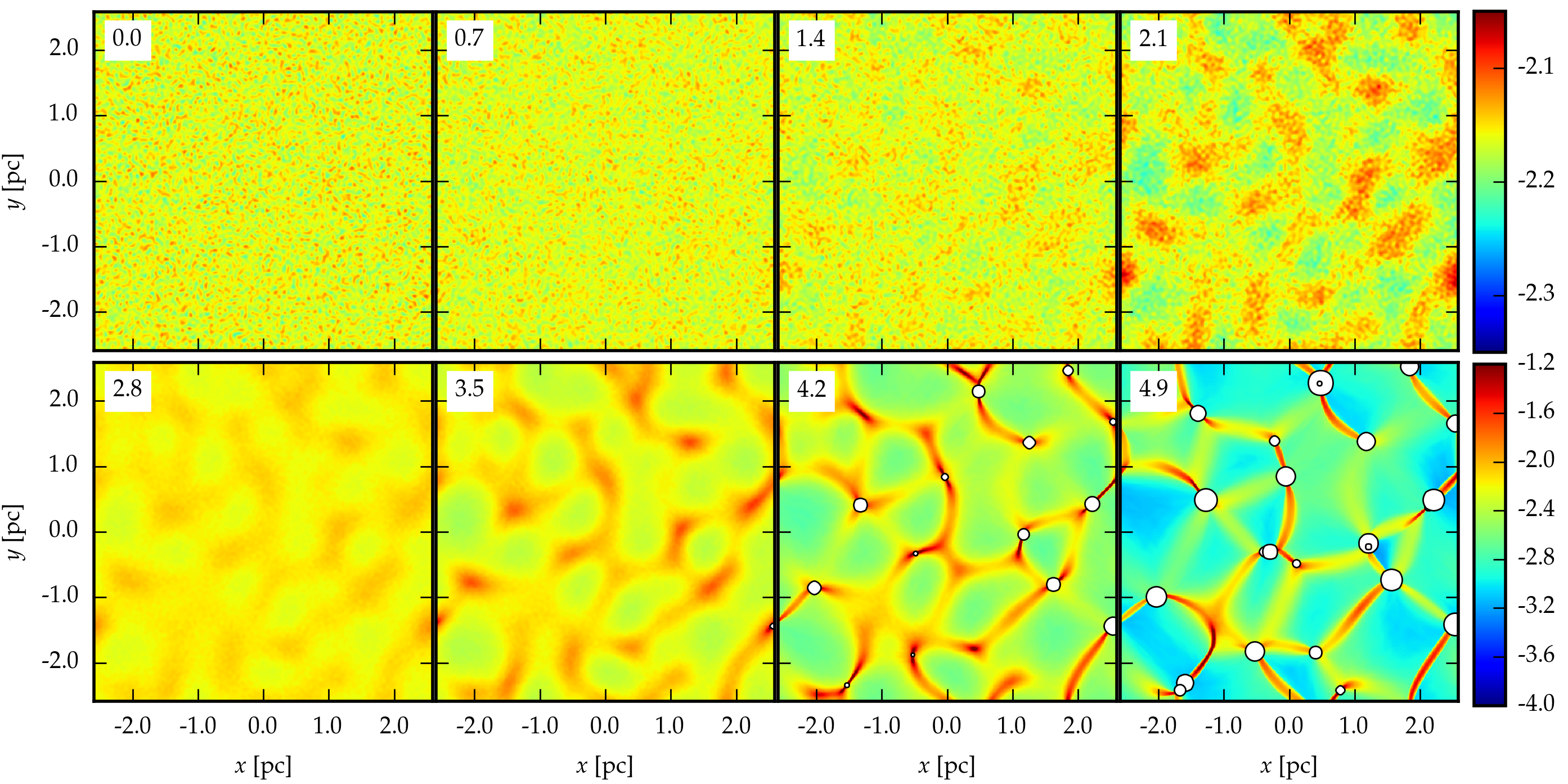}
\caption{Evolution of surface density for the self--gravity dominated layer (model P99). Number at the upper left corner is the time in Myr. Sink particles are plotted
as white circles. Area of a particular circle corresponds to the sink particle mass. Note that the colourscale for upper and bottom row is different.}
\label{fsgr_frag}
\end{figure*}
\else
\fi

\iffigscl
\begin{figure*}
\includegraphics[width=\textwidth]{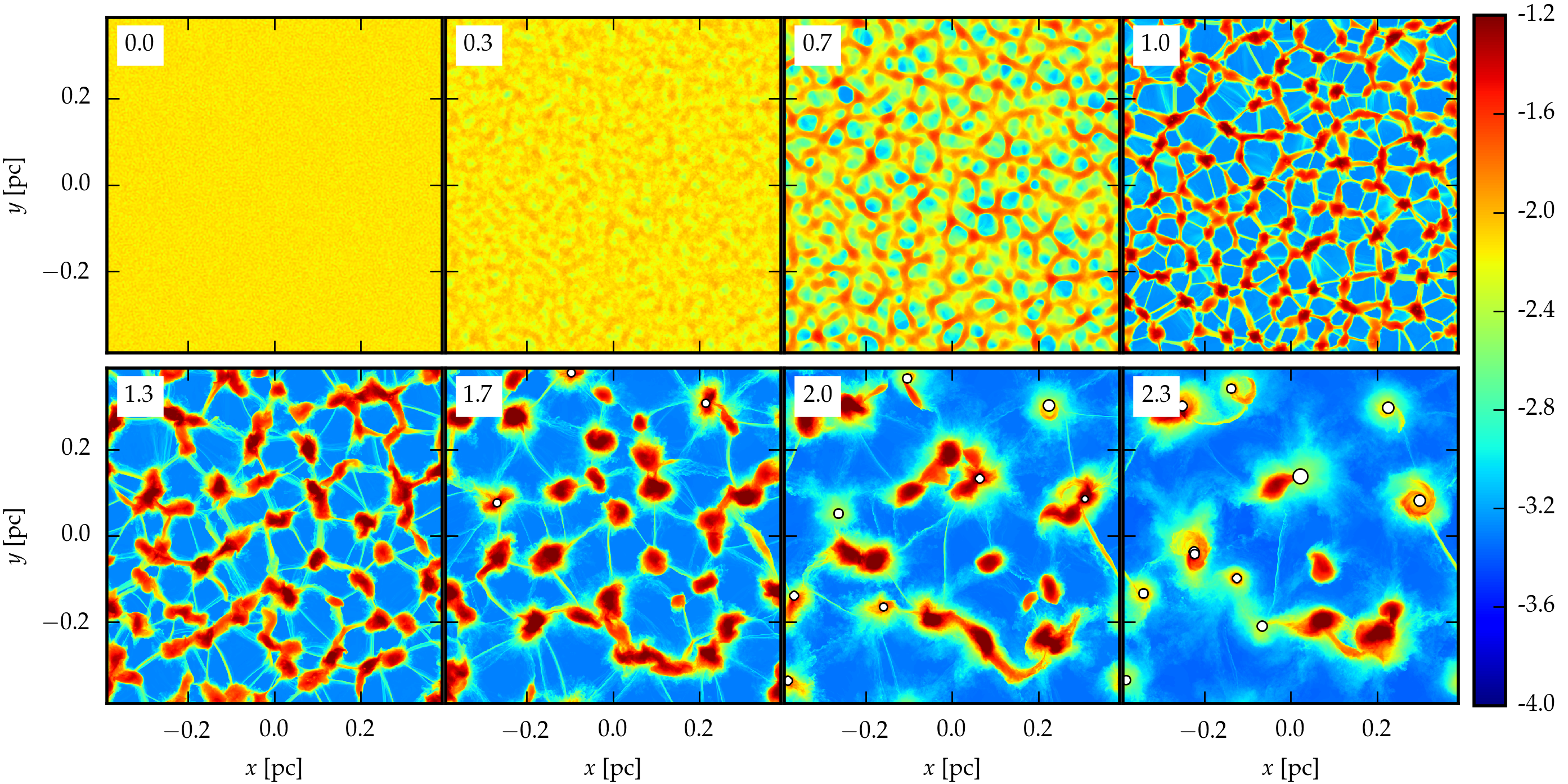}
\caption{Evolution of surface density for the external pressure dominated layer (model P18).
Caption is the same as in Fig. \ref{fsgr_frag}. 
}
\label{fep_frag}
\end{figure*}
\else
\fi

\iflabs
\textit{Fragmenting time scales as function of A}
\else\fi
Since the analytical dispersion relations are often used for estimating fragmenting time--scales and mass of fragments, it is interesting to 
compare these quantities with those found in our simulations. 
We identify gravitationally bound fragments with the algorithm described in Appendix \ref{APP:CLUMPFIND}. 
We experiment with several definitions of fragmenting time $t\FRG$, and conclude that taking 
$t\FRG$ to be an instant when the total mass of gravitationally bound objects exceeds 1/2 of the total mass of the layer is a reasonable estimate for following reasons. 
As formation of bound objects starts, their total mass increases rapidly, so taking another fraction 
of the total mass than 1/2 would not lead to a significantly different time scale.
Time $t\FRG$ is also close to the time $t\SINK$ when sink particles in total contain 1/2 of the total mass of the layer 
(see Table \ref{twn} and left panel of Fig. \ref{ftmcoll_norm}).
%
%


Comparison between $t\FRG$ and E78 estimate $t\EEFOLDFAST = 1/\omega\EFAST$ (\eq{ee78olim}) as a function of parameter $A$ is plotted in the left panel of Fig. \ref{ftmcoll_norm}.
We test the commonly adopted assumption that the fragmentation occurs at a constant number of analytic e--folding times $t\EEFOLDFAST$ regardless of $A$, so we 
multiply $t\EEFOLDFAST$ by a constant to match the data point for run P99. 
The assumption of constant number of e--folding times holds in self--gravity dominated case, but 
it underestimates the collapse time when external pressure dominates.

\iflabs
\textit{dimensional analysis}
\else\fi
The parameters describing the layer are $\Sigma\subO$, $c\subS$ and $P\EXT$. From dimensional analysis it follows that any quantity with the dimension of time is given by
\begin{equation}
t = \frac{c\subS f(A)}{G \Sigma\subO},
\label{edimless_time}
\end{equation}
where $f(A)$ is an unknown function of the parameter $A$.
Since both $t\EEFOLDFAST$ and $t\FRG$ are given by \eq{edimless_time} (with presumably different description of $f(A)$),
the number of e--folding times when fragmentation happens $n\EFOLD = t\FRG/t\EEFOLDFAST$ is only a function of $A$ for any layer.
If we obtain $n\EFOLD(A)$ from our simulations, we can estimate $t\FRG$ for any layer because $t\EEFOLDFAST$ is already known from \eq{ee78olim}.

\iflabs
\textit{efolding time}
\else\fi
The dependence of $n\EFOLD$ on $A$ is shown in the middle panel of Fig. \ref{ftmcoll_norm} (solid line).
Whereas $n\EFOLD$  is nearly constant in the self--gravity dominated case, it strongly increases as the external pressure increases.
Before drawing conclusions from this result, we should verify that it is not simply caused by the fact
that self--gravity dominated models already start with effectively higher initial perturbations.
For any model, the transition to the non--linear regime at time $t\NL$ (when $\Sigma_1 = \Sigma\subO$) occurs
at almost constant number of $t\EEFOLDFAST$ (Table \ref{twn} and middle panel of Fig. \ref{ftmcoll_norm}), 
so the initial perturbations are in this sense comparable among models with different $A$.
Taking $t\NL$ instead of $t = 0$ as the time when the perturbing amplitudes are comparable would
even emphasize the dependence of the number of $t\EEFOLDFAST$ on $A$; when $A$ is near unity, the fragmentation occurs soon after $t\NL$,
when $A$ is small, it takes many e--folding times to reach $t\FRG$.
This behaviour reflects the two different fragmenting scenarios; when $A$ is near unity, the clumps continue collapsing,
when $A$ is small, the clumps are stable and gradually merge, which causes the delay in fragmentation time.
We conclude that neither E78 describes fragmenting time correctly because fragmentation does not
occur at a constant number of e--folding times for various $A$.

\iffigscl
\begin{figure*}
\includegraphics[width=\textwidth]{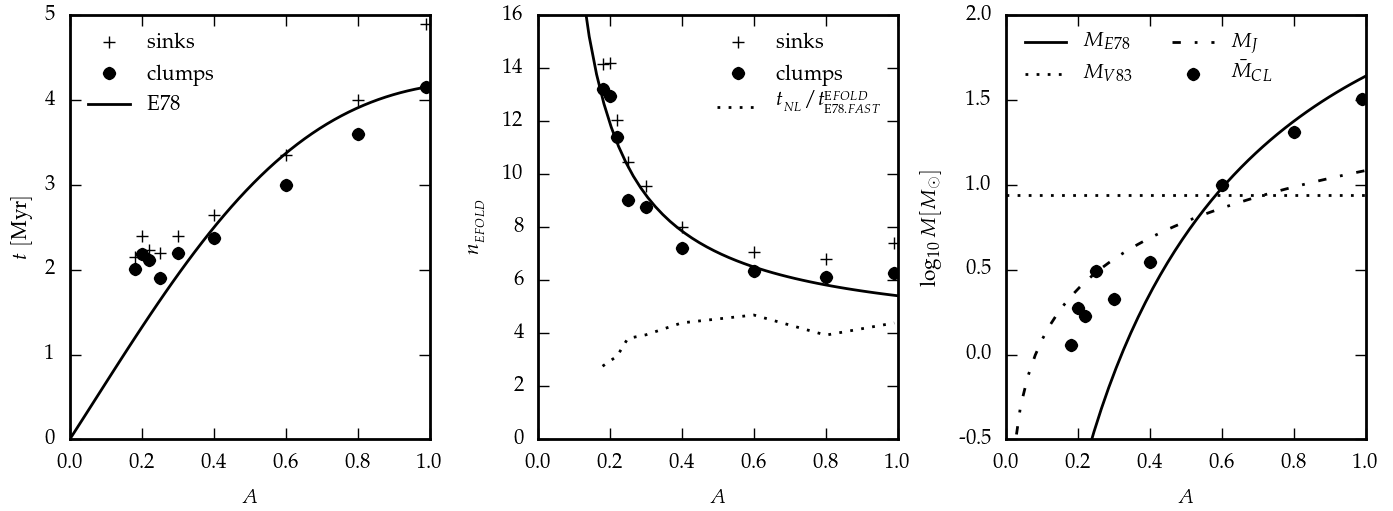}
\caption{Fragmenting time and fragment mass of P models. \figpan{Left panel:} Time when 50\% of the total mass is in the form of sink particles (pluses) and gravitationally bound objects (circles).
The solid line is the analytical estimate for E78, with e--folding time $t\EEFOLDFAST$ scaled so as to match $t\FRG$ for model P99.
\figpan{Middle panel:} Number $n\EFOLD$ of $e$--folding times $t\EEFOLDFAST$ needed to reach time $t\FRG$.
Solid line shows the fit \eq{efit}, dotted line shows the transition between linear and non--linear regime $t\NL/t\EEFOLDFAST$.
\figpan{Right panel:} Average mass of gravitationally bound objects at time $t\FRG$ (circles).
Analytical estimates for E78, V83 and the Jeans mass for the midplane density are shown with solid, dotted and dashed lines, respectively.}
\label{ftmcoll_norm}
\end{figure*}
\else
\fi

\iflabs
\textit{Masses of unstable objects as function of A}
\else\fi
The mean mass of bound objects $\MCL$ at the fragmenting time as a function of $A$ is listed in Table \ref{twn} and shown in the right panel of Figure \ref{ftmcoll_norm}.
Analytical estimates for V83 and E78 ($M_{V83}$ and $M_{E78}$, respectively) and the Jeans mass $M\JEANS$ for the 
midplane density (we use $M\JEANS=4.26 \; A c_s^4/(G^2 \Sigma\subO)$ from eq. (13-33) in \citet{s78}) are plotted by lines.
The fragment masses for E78 and V83 are estimated from the wavelength with the highest growth rate (i.e. $M = \pi \Sigma\subO (\lambda\DRFAST/2)^2$ where $\lambda\DRFAST = 2 \pi / k\DRFAST$).
In the self--gravity dominated case, the mass of fragments is in a good agreement with the E78 prediction. This result is a natural consequence of the
monolithic collapse since the mass of fragments formed in the linear regime does not significantly change later on, during the non--linear collapse.
The thin--shell approximation systematically underestimates the fragment masses for $A \ga 0.6$.

\iflabs
\textit{Masses of unstable objects for low A}
\else\fi
In the external pressure dominated case, the mass of fragments is of the order of the Jeans mass. It is a result of their formation process, since
the fragments are initially subjeans and stable until they assembled approximately the Jeans mass and then collapse.
Therefore, the estimate based on E78 leads to too small fragment mass. \itiitext{In contrast}, V83 overestimates
fragment mass as it does not take into account the decrease of the Jeans mass with increasing external pressure.

\section[Possible pattern formation on layer surfaces]{Layers confined by thermal pressure and possible pattern formation in surface density}

\label{smodint}

\iflabs
\textit{Semi--analytical theory}
\else\fi
\citet{f96} proposes that when fragmentation of an infinitely thin disc becomes non--linear, a triple of modes 
with wavevectors $\|\vec{k}\| \simeq k\DRFAST$ inclined at angles around $60 \degr$ 
(i.e. in the Fourier space forming an equilateral triangle) has the highest growth rate due to interaction.
\citet{wp01} semi--analytically find similar behaviour for the surface of a shell assuming V83. 
As a result, the modes create a hexagonal pattern in surface density.
On the other hand, applying second--order perturbation theory, \citet{mn87a} show that a fragmenting layer breaks into gradually slendering filaments with
no signs of the hexagonal pattern. 

In this section, we test the Fuchs' proposition by searching regular patterns in surface density in our polychromatic simulations. 
Since we find no evidence for any pattern, we perform three dedicated models to test whether pattern formation arises at all under very idealised circumstances.


\iflabs
\textit{Generic names}
\else\fi
Name of the special models exploring possible pattern formation is in the form I${<}A{>}$$\_$${<}N_w{>}$\_[S\textbar D], where the 
first two numbers after "I" represent the parameter $A$ and the number after underscore the number $N_w$ of amplified wavepackets. 
For models with suffix S, the wavepackets are amplified by a smooth function (wavepackets and the function are described in Section \ref{smodintinit}), 
while for models with suffix D, the wavepackets are degenerated to single modes. 

\subsection{Initial conditions for perturbations}

\label{smodintinit}

For the dedicated I models, the amplitudes $\tilde{A}(k)$ of the perturbing wavevectors 
are firstly generated by the same method as for the polychromatic models (see Section \ref{spolyinit} for description).
Then, for models I99\_3\_S and I99\_3\_D, we choose three modes $\vec{k}_i$ ($i = 1,2,3$) with $\|\vec{k}_i\| \simeq k\EFAST$ inclined at an angle around $60^\circ$ 
and amplify amplitudes of all modes inside a circle at centre $\vec{k}_i$ of radius $\dd k$ by a bell shaped curve. 
Thus the perturbing amplitudes $\tilde{A}_n$ are calculated as 
\begin{equation}
\tilde{A}_n(k) =  
\begin{cases}
A_{int} \tilde{A}(k) \exp(-\| \vec{k}, \vec{k}_i \|^2_2/\sigma_A^2) & , \,\, \| \vec{k}, \vec{k}_i \|_2 \leq \dd k \\
\tilde{A}(k) & , \,\, \mathrm{otherwise} \\
\end{cases}
\label{arsc}
\end{equation}
where $\| \vec{k}, \vec{k}_i \|_2 = \sqrt{(k_x - k_{ix})^2 + (k_y - k_{iy})^2}$ and $A_{int}$ and $\sigma_A$ are parameters of the multiplying curve used in
the particular model (Table \ref{tmint}). 
The only parameter differentiating models I99\_3\_D and I99\_3\_S is the radius $\dd k$.
We experimented with amplifying amplitudes of only the selected wavevectors $\vec{k}_i$ (model I99\_3\_D) and 
wavepackets inside non--zero radius $\dd k$ centered on $\vec{k}_i$ (model I99\_3\_S).
The initial conditions for model I99\_1\_D are identical to that of model I99\_3\_D except that only one mode, $\vec{k}_1$ is amplified.

\subsection{Evolution of interacting modes}

\begin{table*}
\begin{tabular}{ccrcccc|ccccc}
Run & $A$ & $n_x \times n_y \times n_z$ & $H\HALF/dz$ & $A_{int}$ & $\sigma_A$ & $dk$ & $i$ & $t\EEFOLDFAST$ & \\
 & & & & & $[\rmn{pc}^{-1}]$ & $ [\rmn{pc}^{-1}]$ & & [Myr] & \\
\hline
I99\_3\_S &  0.99 &    512 $\times$   512 $\times$    32 &   5.1 &   10.0 &   28 &   14 & 1,2,3 &  0.66 \\
I99\_3\_D &  0.99 &    512 $\times$   512 $\times$    32 &   5.1 &   10.0 &   28 &    0 & 1,2,3 &  0.66 \\
I99\_1\_D &  0.99 &    512 $\times$   512 $\times$    32 &   5.1 &   10.0 &   28 &    0 & 1 &  0.66 \\
\end{tabular}
\caption{
\itiitext{Parameters for simulations designed to study mode interaction (models I).} 
First four columns have the same meaning as the columns in Table \ref{tmono}.
Parameters $A_{int}$, $\sigma_A$ and $dk$ characterise properties of the mode amplifying function, \eq{arsc}. 
Further, we list indices $i$ of amplified modes $\vec{k}_i$.
Time $t\EEFOLDFAST$ is as in Table \ref{twn}.}
\label{tmint}
\end{table*}

\iflabs
\textit{Polychromatic models, brief inspection}
\else\fi
Polychromatic models P18 and P99 (i.e. the ones with extreme values of parameter $A$ in our sample) 
enter non--linear regime at time $0.42 \Myr$ and $2.90 \Myr$, respectively (Table \ref{twn}). 
Inspecting corresponding plots in Figs. \ref{fsgr_frag} and \ref{fep_frag} by eye, we do not see pattern formation at any stage of fragmenting process.

\iffigscl
\begin{figure}
\includegraphics[width=\columnwidth]{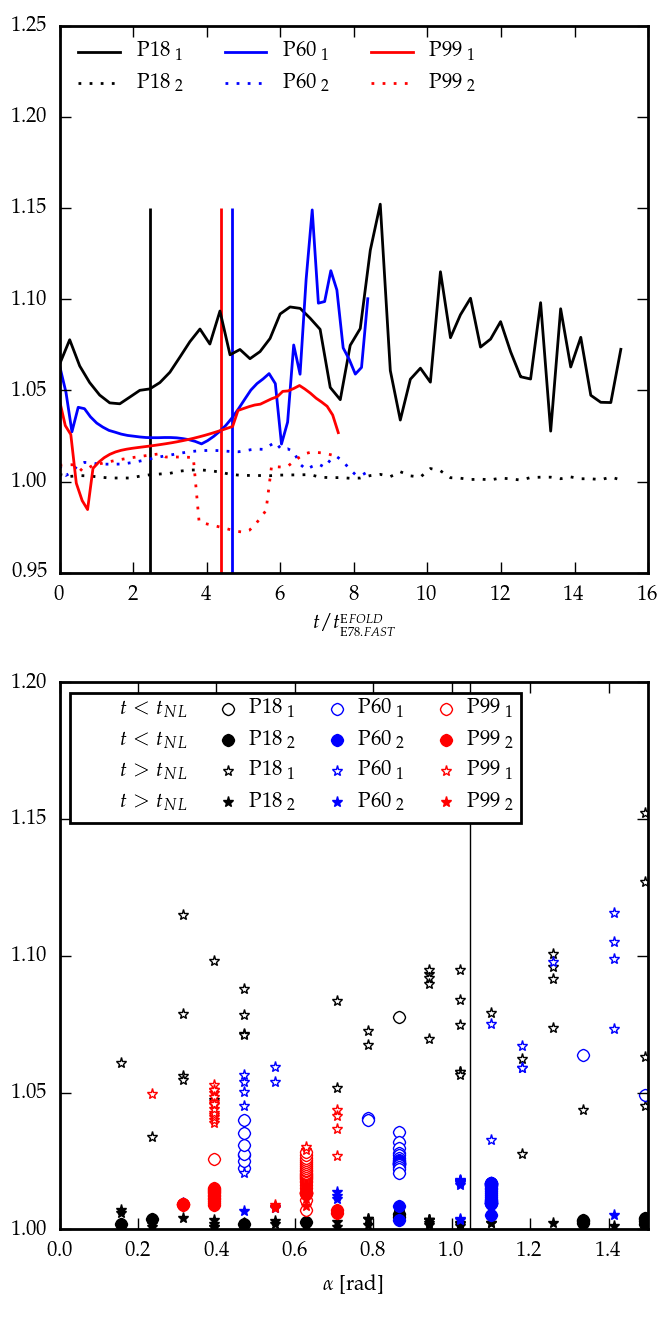}
\caption{\figpan{Upper panel:} Evolution of azimuthal autocorrelation, \eq{ecrosscor}, for models P18 (black line), P60 (blue line) and P99 (red line). 
Subscript characters 1 and 2 refer to two versions of function $B\RAD{ }$; $B\RAD{1}$ and $B\RAD{2}$. 
Vertical bars denote transition from linear to non--linear regime for particular model as represented by its colour.
\figpan{Bottom panel:} Angular separation $\alpha$ between modes with $\CMAX{J}/\CMEAN{J}$ at given time for models P18 (black symbols), P60 (blue symbols) and P99 (red symbols). 
The symbols are circles and asterisks in linear and non--linear regime of fragmentation for each model, respectively. 
Open and filled symbols represent analysis with $B\RAD{1}$ and $B\RAD{2}$, respectively. 
The angle 60\degr, which is expected to be adjoined by the triple of the highest growing modes in scenario proposed by \citet{f96}, is denoted 
by the vertical solid line.
}
\label{fmode_int_p}
\end{figure}
\else
\fi

\iffigscl
\begin{figure}
\includegraphics[width=\columnwidth]{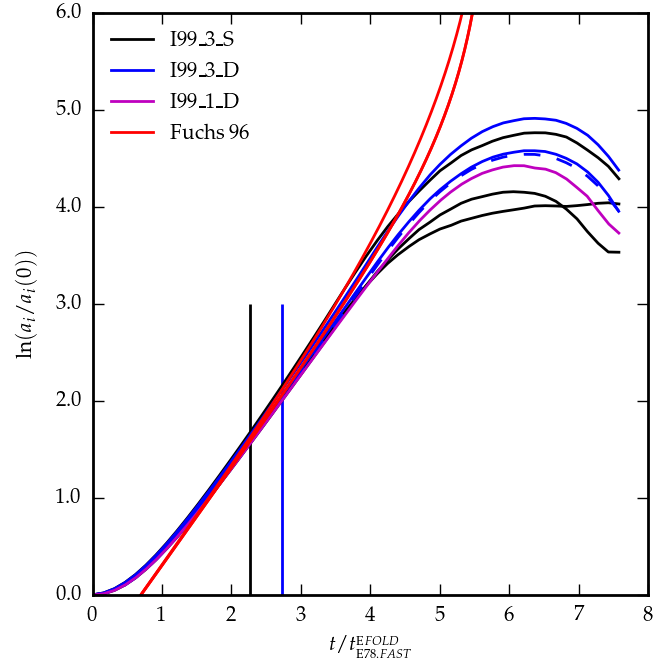}
\caption{
Amplitude evolution of the three amplified modes for models I99\_3\_S (black line) and I99\_3\_D (blue line), and for 
the single amplified mode for model I99\_1\_D (magenta line). 
The mode of model I99\_3\_D which is identical to the amplified mode of model I99\_1\_D is shown by the dashed blue line. 
Numerical solution to the second order perturbation equations proposed by \citet{f96} is plotted by red lines.
The transition from the linear to non--linear regime is shown by vertical bars. 
}
\label{fmode_int_i}
\end{figure}
\else
\fi

\iflabs
\textit{Crosscorelation study -- definition}
\else\fi
We perform more rigorous analysis to find a firmer support for this statement.
Let Fourier transform of the surface density be $B(k_x, k_y)$.
A regular pattern in the surface density would manifest itself as an anisotropy of function $B$.
We look at the plane $(k_x, k_y)$ as it is given in polar coordinates specified by radius $\|\vec{k}\|$ and azimuthal angle $\alpha$.
To search for the anisotropy, we analyse function $B$ evaluated on two different radially-averaged subsets in the $(k_x, k_y)$ plane.
In the first case, $B$ is evaluated on a thin annulus with radius $\|\vec{k}\| = k\EFAST$,
i.e. $B\RAD{1}(\alpha) \equiv B(\|\vec{k}\| = k\EFAST, \alpha)$. 
In the second case, the annulus is wider so it includes more modes around radius $k\EFAST$, 
i.e. $B\RAD{2}(\alpha) \equiv B(k\EFAST/2 \leq \|\vec{k}\| \leq 3k\EFAST/2, \alpha)$. 
If two modes inclined at an angle $\tilde{\alpha}$ have the highest amplitudes, function $B\RAD{J}(\alpha)$ (letter $J$ stands for 1 or 2) 
has maxima separated by the angle $\tilde{\alpha}$. 
Thus the amplitudes of modes adjoined by the angle $\alpha$ are proportional to 
the value of azimuthal autocorrelation of $B\RAD{J}(\alpha)$, i.e.
\begin{equation}
c_J(\alpha) = \int_0^\pi B\RAD{J}(\alpha') B\RAD{J}(\alpha + \alpha') \dd \alpha'.
\label{ecrosscor}
\end{equation}
Function $c_J(\alpha)$ detects interaction of modes inclined by any angle, not necessarily $60 \degr$. 

When analysing the data, it is important to note that the autocorrelation function always attains its maximum at zero. 
The maxima due to two modes adjoined by an angle are local maxima of function $c_J$. 
We denote $\CMAX{J}$ the highest local maximum of function $c_J$
after its global maximum $c_J(0)$. 
During the simulations, the value of $c_J$ increases as amplitudes of modes increase. 
Since the interacting modes are predicted to grow faster than the rest of modes, 
we normalise $\CMAX{J}$ with the mean $\CMEAN{J}$.

\iflabs
\textit{Crosscorelation study -- results}
\else\fi
Time dependence of $\CMAX{1}/\CMEAN{J}$ and $\CMAX{2}/\CMEAN{2}$ for models P18, P60 and P99 is shown in the upper panel of Figure \ref{fmode_int_p}.
Vertical bars indicate the transition to the non--linear regime. 
We detect no significant growth of $\CMAX{J}/\CMEAN{J}$ after the models enter the non--linear regime indicating 
isotropy in the plane $B(k_x, k_y)$ and thus no formation of a regular pattern. 

The bottom panel of Fig. \ref{fmode_int_p} shows the angle $\alpha$ at which $\CMAX{J}/\CMEAN{J}$ is attained. 
Data points representing linear and non--linear regime are distinguished by circles and asterisks, respectively. 
In the linear regime, there is no preferred angle adjoined by modes with the highest growth rate.
This result is not surprising since the modes are predicted to grow independently on one another in the linear regime.
If modes started interacting in the non--linear regime, asterisks would be clustered around a particular angle 
($\alpha = 60 \degr$ is represented by a vertical line) 
with concomitant increase of $\CMAX{J}/\CMEAN{J}$. 
Instead, the symbols show neither preferential clustering nor a rapid increase
of $\CMAX{J}/\CMEAN{J}$.
The absence of both signposts strongly disfavours the possibility of regular pattern formation.

\iflabs
\textit{Growth rate of the selected modes}
\else\fi
Models I enable us to study the evolution of amplitudes $a_i$ of the individual modes $\vec{k}_i$; this is plotted in Fig. \ref{fmode_int_i}.
Models I99\_3\_S (black lines) and I99\_3\_D (blue lines) contain the triple of modes, which according to \citet{f96}, should 
have an increased growth rate in the non--linear regime. 
Red lines show numerical solution to the second order perturbation equations for the thin shell approximation (cf. eqs. (28) in \citep{f96}) with 
the initial mode amplitudes of model I99\_3\_D
\footnote{
Since the initial conditions for models I are not eigenfunctions, they do not exhibit the constant growth rate from the beginning. 
Eqs. (28) in \citep{f96} are derived for eigenfunctions, so they already starts with the constant growth rate. 
To compensate for the offset, we divide the latter by a constant so that their amplitudes equal to that of the former at the end of the linear regime.
}
.The vertical bars mark the transition from the linear to non--linear regime. 
Recall that the growth rate $\omega$ is the time derivative of the plotted functions. 
The growth rate obtained from our simulations does not follow the increased growth rate predicted by Fuchs. 
In contrast, the simulated growth rate even decreases in the non--linear regime.

In the previous paragraph, we demonstrate that the possible interaction between the triple of modes does not increase their growth rate above $\omega \EFAST$. 
Do the triple of modes interact at least to some degree? 
To assess this question, we use the same initial conditions as for model I99\_3\_D, but we amplify only one mode, $\vec{k}_1$ (model I99\_1\_D). 
Amplitude of this mode (magenta line in Fig. \ref{fmode_int_i}) evolves very close to that of the corresponding mode in model I99\_3\_D (blue dashed line). 
Moreover, the similar shape between these two curves suggests that the non--linear terms due to the two other modes are not important.

%

\iflabs
\textit{Notes about our approach}
\else\fi
Note that to study pattern formation, we investigate mode interaction only in the azimuthal direction $\alpha$ of the Fourier space.
We do not investigate mode interaction in radial direction $\|\vec{k}\|$, which
apparently arises in the non--linear regime.
Our findings do not contradict the mechanism proposed by \citet{mn87a, mn87b}, which leads to randomly oriented filaments.

%


\section[Accreting layers]{Layers accreting from one side, and contained by thermal pressure on the other}

\label{saccreting}

\iflabs
\textit{Overview of the section}
\else\fi
In this section, we study layers 
accreting homogeneous medium from the upper surface and bounded by thermal pressure from the lower surface. 
We investigate their dispersion relation and subsequent evolution in the non--linear regime.
These models approximate a part of a shell sweeping up the ambient medium on one surface and backed by thermal pressure on the other. 
As we show in Section \ref{scol_time} that shells around \HII regions fragment in the pressure dominated case, 
we start the simulations in this case, setting $A = 0.3$ at the beginning of a simulation.


\iflabs
\textit{Generic names}
\else\fi
The generic name of accreting models is in the form A${<}A{>}$$\_$${<}\mathscr{M}{>}$$\_{<}T\AMB{>}\_[N$\textbar$L]$. 
First number after letter A represents the value of the parameter $A$ at the beginning, followed by the Mach number $\mathscr{M}$ of 
the accreting medium and the temperature $T\AMB$ of the ambient medium imposing the thermal pressure. 
The suffix L or N indicates whether the simulation terminates in linear or non--linear regime, respectively.
\itiitext{For example, simulation A30\_08\_10\_L treats a layer with $A = 0.30$, which accretes at velocity $\mathscr{M} = 8$, and which is 
backed with an ambient medium of temperature 10 000 K.}

We use two kinds of models to address two different issues.
Models with suffix L are used for studying evolution in the linear regime (the dispersion relation and flows inside the layer).
For this purpose, it is sufficient to use computational domain in $xy$ directions significantly ($8 \times$) smaller than in corresponding
non--accreting model P30. Consequently, we can afford to use two times higher resolution $H\HALF/ \dd z$, even though a number of grid cells in $x$ and $y$ directions are $4$ times smaller.
Models with suffix N focus on evolution in the non--linear regime, where the priority is to include many Jeans masses inside the computational domain.
For this purpose, we use the same size of the computational domain  and the resolution as for model P30.


\subsection{Initial conditions for perturbations}


\label{saccinit}

Initial perturbations are generated by the same method as for the polychromatic models (Section \ref{spolyinit}). 

\subsection{Evolution in linear regime}

\label{saccretinglin}

\begin{table*}
\begin{tabular}{ccrcccc|ccccccc}
Run & $A$ & $n_x \times n_y \times n_z$ & $H\HALF/dz$ & $\mathscr{M}$ & $\rho\ACC$ & $T\AMB$ & $t\EEFOLDFAST$ & $\ATO$ & $\ATT$ & $\SIGTO/\Sigma\subO$ & $\SIGTT/\Sigma\subO$ & &  \\
 &  &  &  &  & $[10^{-21} \rmn{g.cm}^{-3}]$ & [$\rmn{K}$] & & & & &  \\
\hline
A30\_08\_10\_L &  0.30 &   128 $\times$   128 $\times$    64 &   7.8 &     8 &  1.660  & 10000 & 0.25 &  0.39 &  0.47 &  1.37 &  1.76 \\
A30\_20\_10\_L &  0.30 &   128 $\times$   128 $\times$    64 &   7.8 &    20 &  0.320  & 10000 & 0.25 &  0.33 &  0.38 &  1.16 &  1.35 \\
A30\_50\_10\_L &  0.30 &   128 $\times$   128 $\times$    64 &   7.8 &    50 &  0.052  & 10000 & 0.25 &  0.31 &  0.33 &  1.06 &  1.15 \\
A30\_08\_03\_L &  0.30 &   128 $\times$   128 $\times$    64 &   7.8 &     8 &  1.660  &   300 & 0.25 &  0.39 &  0.48 &  1.39 &  1.78 \\
A30\_20\_03\_L &  0.30 &   128 $\times$   128 $\times$    64 &   7.8 &    20 &  0.320  &   300 & 0.25 &  0.34 &  0.41 &  1.17 &  1.48 \\
A30\_50\_03\_L &  0.30 &   128 $\times$   128 $\times$    64 &   7.8 &    50 &  0.052  &   300 & 0.25 &  0.31 &  0.39 &  1.07 &  1.39 \\
\end{tabular}
\caption{Parameters of accreting simulations intended to study evolution in the linear regime (models A). 
First four columns have the same meaning as the columns in Table \ref{tmono} (parameter $A$ is taken at time zero).
Further we list Mach number $\mathscr{M}$ and density $\rho\ACC$ of the accreting medium, temperature of the ambient medium $T\AMB$, and surface density 
and the value of parameter $A$ at time $t\EEFOLDFAST$ and $2t\EEFOLDFAST$. Surface density at the beginning is denoted $\Sigma\subO$.
}
\label{tal}
\end{table*}

\iflabs
\textit{Fragmentation in linear regime}
\else\fi
Numerically obtained dispersion relations in time interval $(t^{fit}\subO, t^{fit}_1) = (1/\omega\EFAST, 2/\omega\EFAST)$ 
for models with $T\AMB = 300 \rmn{K}$, i.e. A30\_08\_03\_L, A30\_20\_03\_L and A30\_50\_03\_L (Table \ref{tal}) are plotted in the upper panel of Figure \ref{facc_lin}. 
As the layer accretes, its surface density and parameter $A$ increase, we list these quantities at $t^{fit}\subO$ and $t^{fit}_1$ in the last four columns of Table \ref{tal}.
We compare the results with E78 for two extreme values of parameter $A$ attained in the simulations; 
one corresponds to $t=0$, i.e. $A=0.30$, the other to the model with the highest $A$ at $t^{fit}_1$, 
i.e. $A=0.48$.
If the dispersion relation for the accreting layer were the same as for the thermal--pressure confined layer, all models would fit between these curves.
However, the range of unstable wavenumbers extends to significantly higher values of $kH\HALF$. 
The highest growth rate is around factor 2 higher than that of E78.
To study possible influence of the temperature in the ambient medium, we calculate the same models with temperature $T\AMB =10^4 \rmn{K}$ 
(lower panel of Fig. \ref{facc_lin}).
Both kinds of models reproduce similar features indicating that the dispersion relation depends only weakly on the temperature 
of the ambient medium.
Note that these features are unlikely to be caused by resolution, which is here higher (7.8 cells per $H\HALF$; Table \ref{tal}) than that for 
polychromatic models ($\simeq 4 H\HALF$; Table \ref{twn}), which reproduce their analytic dispersion relation very well.

\iflabs
\textit{Circular motions inside layer}
\else\fi

Evolution of density and velocity field inside the accreting layer (model A30\_08\_10\_L) is shown in Figure \ref{fslice_m08}.
As the upper part of the layer is mildly corrugated, the different directions of incidence between the ram pressure and 
thermal pressure lead to a dynamical instability similar to the one
described by \citet{v83}, and it results in gas motions towards the convex parts (panel (a)). The gas then moves through the layer and protrudes from contact discontinuity until it 
is eventually stopped by gravitational attraction of the layer (panels (b) and (c)). 
Formation of the protrusions at the contact discontinuity broadens the unstable range of the dispersion relation towards higher $kH\HALF$ as is apparent from Fig. \ref{facc_lin}.
At the same time, the corrugations of the bottom surface of the layer shield the hot gas in concave regions, leading to a pressure drop and thus 
cause flows of hot gas toward the concave regions. Upward flows of hot gas cause upward motions of cold gas inside the layer (panels (b) and (c)). 
However, the different nature of boundary conditions leads to different behaviour of the flows when they reach the surface.
Whereas the upper flows are eroded by a shock, the downward flows can significantly corrugate the contact discontinuity (panel (d)) because this surface is pliable.
The self--gravity then pulls the protruding gas, while it is still replenished from the
original direction, resulting in circular motions (panel (d)).
At this time ($\omega\EFAST t \simeq 4$), the growth of perturbations with $kH\HALF \ga 0.7$ is saturated, so 
the range of unstable wavenumbers is comparable to that of the thermal--pressure confined layer. 

\iffigscl
\begin{figure}
\includegraphics[width=\columnwidth]{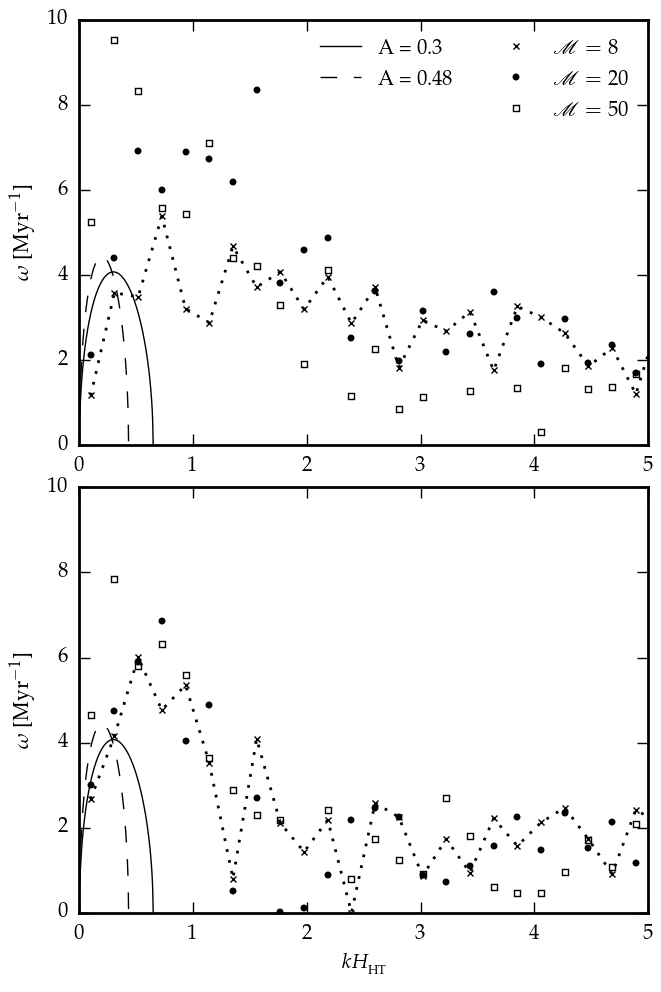}
\caption{Dispersion relations for layers with $A=0.3$ accreting homogeneous medium at Mach numbers 8, 20 and 50 (markers). 
Solid and dashed lines show E78 dispersion relation for two extreme values of $A$; $A = 0.3$ and $A = 0.48$, respectively.
\figpan{Upper panel:} temperature of the ambient medium is 300K (models A30\_08\_03\_L, A30\_20\_03\_L and A30\_50\_03\_L).
The dotted line is to guide the eye for $\mathscr{M}$ = 8 model. 
\figpan{Bottom panel:} temperature of the ambient medium is $10^4$K (models A30\_08\_10\_L, A30\_20\_10\_L and A30\_50\_10\_L).}
\label{facc_lin}
\end{figure}
\else
\fi

\iffigscl
\begin{figure}
\includegraphics[width=\columnwidth]{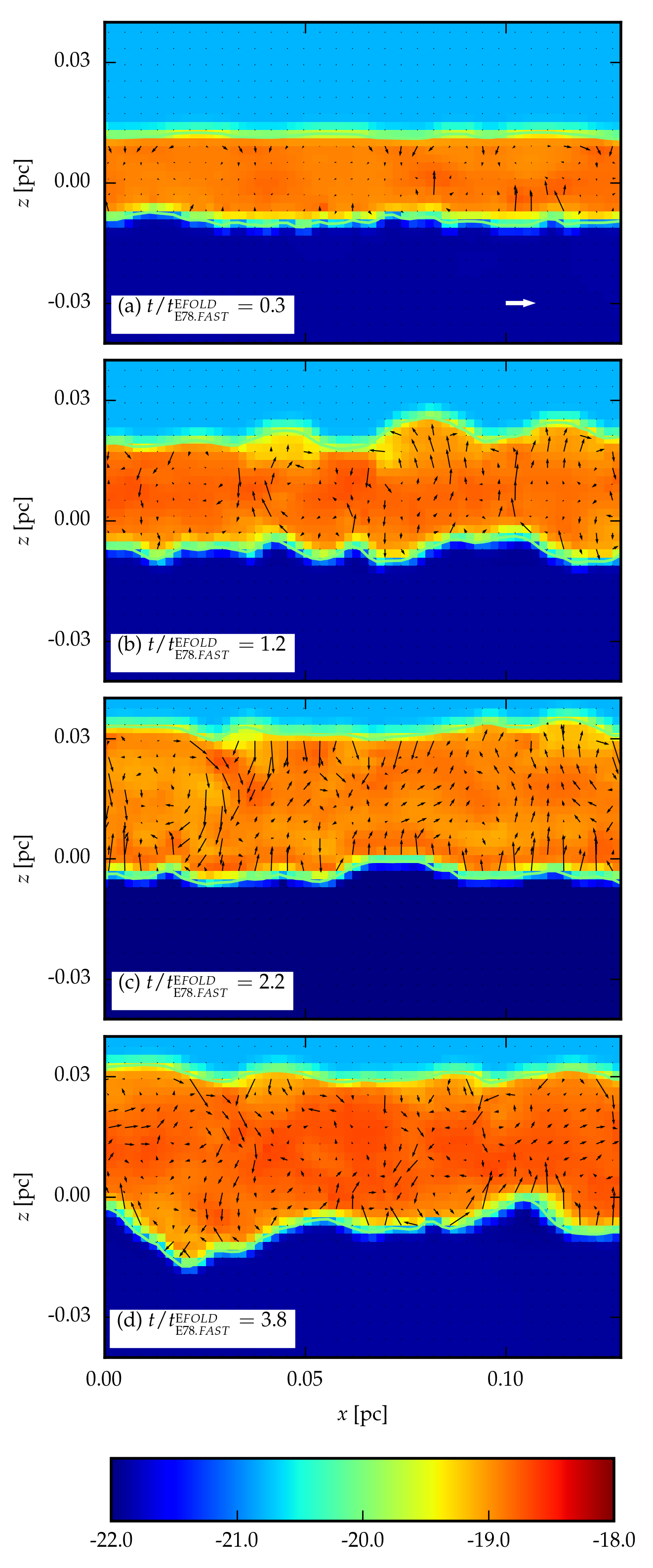}
\caption{Density profile and velocity field on cross sections for model A30\_08\_10\_L. The gas is accreted from the top. 
The velocity vectors are plotted only for the gas inside the layer to avoid confusion
The length of the thick white arrow in panel (a) corresponds to velocity 1 Mach number inside the layer.
}
\label{fslice_m08}
\end{figure}
\else
\fi

\subsection{Evolution in non--linear regime}

\label{saccnl}

\begin{table*}
\begin{tabular}{ccrcccc|cccccccc}
Run & $A$ & $n_x \times n_y \times n_z$ & $\frac{H\HALF}{\dd z}$ & $\mathscr{M}$ & $\rho\ACC$ & $\dot{\Sigma}$ & $N\JEANS$ & $M\JEANS$ & $\MCL$ & $t\EEFOLDFAST$ & $t\FRG$ & $A\FRG$ & $\frac{\Sigma\FRG}{\Sigma\subO}$  \\
 &  &  &  & & [$10^{-21} \rmn{g.cm}^{-3}$] & $\frac{\Msun}{\Myr \Pc^2}$ &  &  [$\Msun$] & [$\Msun$] & [Myr] & [Myr] & &  \\
\hline
A30\_05\_03\_N & 0.30 &   512 $\times$   512 $\times$    64 &   3.9 &     5 & 4.94 &   76 &   9.8 &   3.6 &   4.0 &  0.25 &  1.5 &  0.73 &   3.5 \\
A30\_10\_03\_N & 0.30 &   512 $\times$   512 $\times$    64 &   3.9 &    10 & 1.27 &   39 &   9.8 &   3.6 &   3.5 &  0.25 &  1.7 &  0.59 &   2.4 \\
A30\_20\_03\_N & 0.30 &   512 $\times$   512 $\times$    64 &   3.9 &    20 & 0.32 &   20 &   9.8 &   3.6 &   3.4 &  0.25 &  1.8 &  0.59 &   2.4 \\
\end{tabular}
\caption{Parameters of accreting simulations intended for the non--linear regime of fragmentation (models A).
First four columns have the same meaning as the columns in Table \ref{tmono} and 
quantities $N\JEANS$, $M\JEANS$, $\MCL$,  $t\EFAST$ and $t\FRG$ are explained in Table \ref{twn}. 
We further list the Mach number $\mathscr{M}$ and density $\rho\ACC$ of the accreting medium, accreting rate $\dot{\Sigma}$, 
and values of parameter $A$ and surface density at the time $t\FRG$ ($\Sigma\subO$ is the surface density at the beginning).}
\label{tan}
\end{table*}

\iflabs
\textit{Nonlinear fragmentation for selfgrav. layer}
\else\fi
The subsequent non--linear fragmentation is investigated by models A30\_05\_03\_N, A30\_10\_03\_N and A30\_20\_03\_N (Table \ref{tan}).
Evolution of surface density is shown in Figure \ref{facc_frag}.
Model A30\_05\_03\_N accretes at the highest rate (cf. the 6th column in Table \ref{tan}) and substantial part of fragmentation occurs in self--gravity dominated case ($A \ga 0.6$; 
value of $A$ is at the upper right corner of each frame of Fig. \ref{facc_frag}).
Fragments emerge at $t = 1 \rm{Myr}$ and subsequently collapse, forming filaments and then sink particles at their junctions ($t = 1.5 \rm{Myr}$ and $t = 2 \rm{Myr}$).
Although the boundary conditions at the upper surface are different from that of purely thermal pressure confined layers, 
we see a similar evolutionary pattern as for the monolithic collapse (model P99 described in Section \ref{snonlinear}).

\iflabs
\textit{Nonlin frag. for pressure dominated layer}
\else\fi
\itiitext{In contrast}, the amount of gas accreted in model A30\_20\_03\_N is substantially smaller and the main part of fragmentation is 
accomplished still in pressure--dominated case. 
At the beginning of the fragmentation ($t \la 1.0 \rm{Myr}$), the density inside the layer is approximately constant and 
the overdensities seen in panels at $t = 0.5 \rm{Myr}$ and $t = 1.0 \rm{Myr}$ are corrugations of the contact discontinuity.
The corrugations then gradually merge until they become gravitationally unstable and collapse (panel at $t = 2.0 \rm{Myr}$). 
Note that the values of $A$ seen in Fig. \ref{facc_frag} are slightly higher than due to the accretion since cooling of hot gas at the contact discontinuity 
also contributes to the surface density, which increases $A$. 
The gradual merging forms less defined and more fluffy filaments than the direct collapse at smaller Mach numbers.
Thus the fragmenting process of model A30\_20\_03\_N is very similar to the coalescence driven collapse seen in model P18.

\iflabs
\textit{Characteristic masses}
\else\fi
The mean clump mass $\MCL$ for models A30\_05\_03\_N, A30\_10\_03\_N and A30\_20\_03\_N ($3.4 \Msun$ to $4.0 \Msun$, Table \ref{tan}) is close to 
the Jeans mass ($3.6 \Msun$ for $A=0.3$).
We interpret values of $\MCL$ as follows: As the layer starts accreting in pressure dominated case, 
its density (according to \eq{erho0}) and therefore the Jeans mass $M\JEANS$ is only a weak function of $A$.
The highest change of $M\JEANS$ is for the model with the highest accreting rate, i.e. A30\_05\_03\_N; $M\JEANS$ decreases only by factor $1.7\times$ from beginning to the fragmenting time.
Compared to non--accreting model P99, where $\MCL = 2.1 \Msun$ (Table \ref{twn}), the accreting simulations have at most by factor 2 higher $\MCL$.
Therefore the Jeans mass at $t = 0$ is a relatively good estimate for fragment mass also for accreting layers, so we can use it as a proxy when estimating 
fragments properties for an accreting shell in Section \ref{scol_time}.

\iflabs
\textit{Validity of the models}
\else\fi
Finally, we discuss the influence of our cooling method (described in Section \ref{smab}) on the boundary conditions.
Recall that intermixing the warm gas into the layer and its imminent cooling deplete the warm gas 
on the lower side of the layer. The depleted gas is compensated by warm gas inflow, which causes ram pressure on the contact discontinuity.
If the ram pressure is strong enough, it would change the contact discontinuity to a shock front which is an artificial and unwanted behaviour 
(e.g. surface gravity waves would be suppressed), so we should check that this does not occur.
From the beginning to the fragmenting time, the highest ratio between the ram and thermal pressure at the contact discontinuity is 0.03, 0.12 and 0.5 
for models A30\_05\_03\_N, A30\_10\_03\_N and A30\_20\_03\_N, respectively. Thus the ram pressure is negligible when compared to the thermal pressure 
for models A30\_05\_03\_N, A30\_10\_03\_N and it is not dominant for model A30\_20\_03\_N so that the contact discontinuity is preserved in all simulations.


\iffigscl
\begin{figure*}
\includegraphics[width=\textwidth]{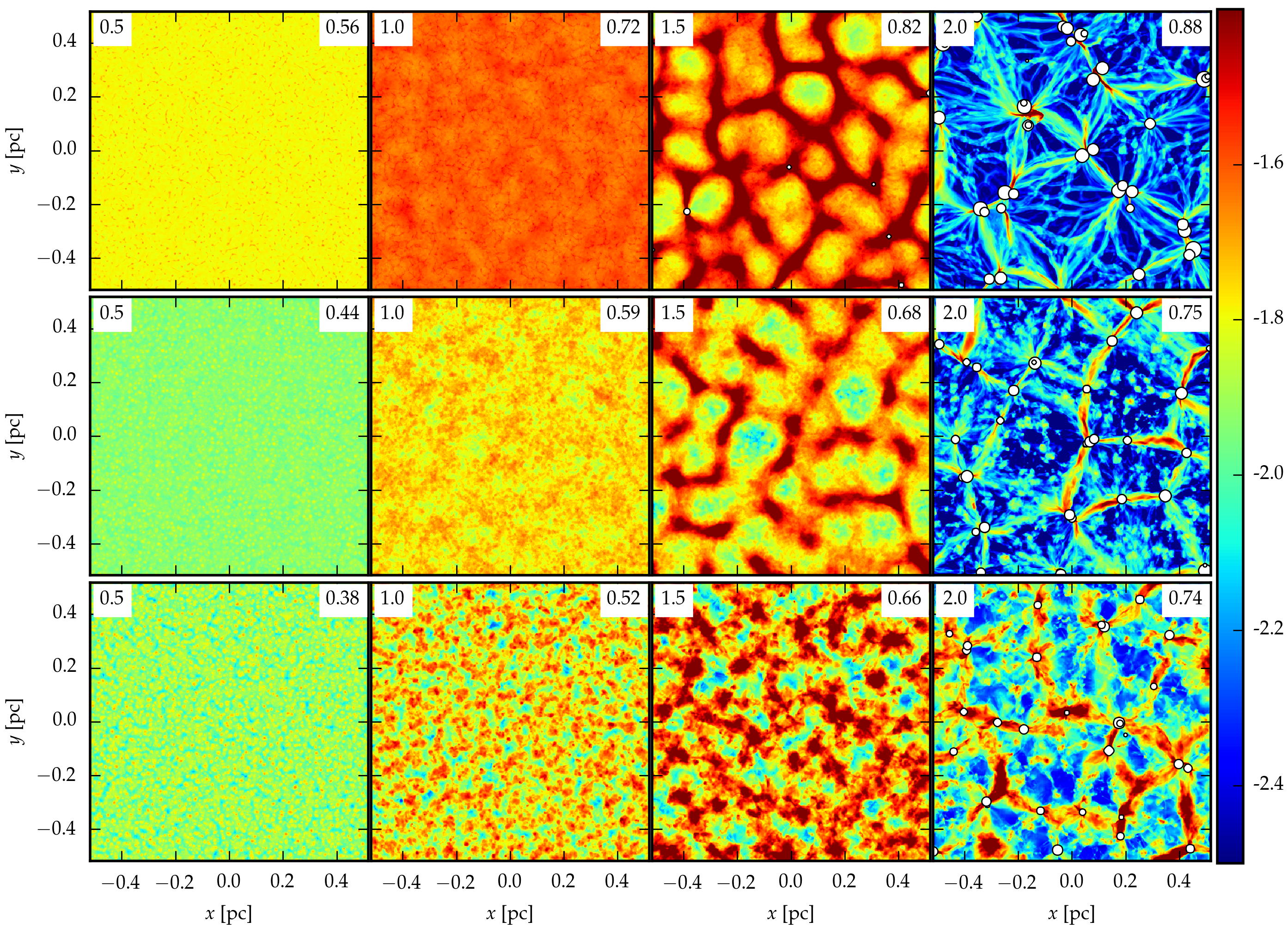}
\caption{Evolution of surface density for accreting simulations. 
\figpan{Top row:} $\mathscr{M}$ = 5 (model A30\_05\_03\_N), \figpan{Middle:} $\mathscr{M}$ = 10 (A30\_10\_03\_N), \figpan{Bottom:} $\mathscr{M}$ = 20 (A30\_20\_03\_N).
Time in Myr is in the upper left corner, the value of parameter $A$ in the upper right corner.} 
\label{facc_frag}
\end{figure*}
\else
\fi



\section{Fragmentation of a shell swept up by an expanding \HII region}

\label{scol_time}

\iflabs
\textit{Assumptions, their justification}
\else\fi
We use results of previous sections to estimate properties of a fragmenting 
shell (e.g. fragmenting time, mass) swept up around an \HII region. 
The \HII region is powered by a massive star emitting $\dot{N}$ ionising photons per second situated in a homogeneous 
medium with density $\rho\ACC$ (we use subscript ACC for the surrounding medium 
since it is accreted in the reference frame of the shell). 
For convenience, we define $\dot{N}_{49} = [\dot{N} / 10^{49 }\rmn{s^{-1}}]$, 
$c_{s2} = [c\subS/0.2 \rmn{km/s}]$, and 
$n_{3} = [n\ACC /10^3 \rmn{cm^{-3}}]$ where $c\subS$ is sound velocity in the swept up shell, and $n\ACC$ is the number density in the surrounding medium.
Following \citet{wb94a}, we assume that fragmentation is accomplished in pressure dominated case, 
which means that the shell fragments by coalescence driven collapse.
At the end we show that our results are consistent with this assumption.

\subsection{An analytic estimate for fragmentation of a shell swept up by an expanding \HII region}

\iflabs
\textit{Move from accreting to nonaccreting models}
\else\fi
The remarkable property of our accreting simulations is that their fragmenting time depends very weakly on
the Mach number (Table \ref{tan}) and differs at most by factor 1.5 from their non--accreting counterpart (model P30). 
Mass of the fragments also differs at most by factor two from the non--accreting model. This behaviour suggests that 
fragmenting time and mass of fragments of a non--accreting layer can be used to estimate corresponding quantities for an accreting layer.

\iflabs
\textit{Introduction of modified fragmenting integral}
\else\fi
Further we assume that a thermal pressure--confined layer with time dependent surface density $\Sigma (t)$ and external pressure $P\EXT (t)$ fragments at any instant 
as the same layer with constant $\Sigma\subO$ and $P\EXT$. Thus, when fragmentation starts at time $t\subO$, it is accomplished at time $t_1$ fulfilling
\begin{equation}
1 = \int_{t\subO}^{t_1} \frac{\omega\EFAST (t)}{n\EFOLD (A(t))} \dd t = \int_{t\subO}^{t_1} \frac{\sqrt{0.276} \pi G \Sigma (A(t))}{c\subS A(t) n\EFOLD (A(t))} \dd t,
\label{efrg_integral}
\end{equation}
where the second equality follows from \eq{ee78olim}.
Fitting measured data (Table \ref{twn} and middle panel in Fig. \ref{ftmcoll_norm}), we obtain 
\begin{equation}
n\EFOLD = 1.63/A + 3.79.
\label{efit}
\end{equation}

\iflabs
\textit{Comparison for a planar layer}
\else\fi
We use our accreting models (with time dependent $\Sigma$) to test the assumption made in the previous paragraph.
Models A30\_05\_03\_N, A30\_10\_03\_N and A30\_20\_03\_N fragment at time 1.5 (1.36) Myr, 1.7 (1.64) Myr and 1.8 (1.85) Myr, respectively, where 
the first number is obtained from the simulation and the bracketed number is from \eq{efrg_integral} showing a good agreement.

\iflabs
\textit{Form of the fragmenting integral for Spitzer solution}
\else\fi
The radius $R$ of a shell around an \HII region expands according to $R = R\ST (1 + 7c_{II} t/4R\ST)^{4/7}$ \citep{s78}, where $R\ST$ and 
$c_{II}$ is Str\"{o}mgren radius and sound velocity in the ionised gas, respectively. 
Surface density of the shell and ram pressure due to swept up gas evolve as $\Sigma = \rho\ACC R(t)/3$ and $P\RAM = \rho\ACC (\dd R/\dd t)^2$. 
Thermal pressure $P\EXT$ inside the \HII region is equal to $P\RAM$. 
Thus approximating a small patch on the shell wall as a layer, its parameter $A$ (\eq{a}) evolves as 
\begin{equation}
A = \left \{ 1 + \frac{18 c_{II}^2 }{\pi G \rho\ACC (R\ST(1 + 7c_{II} t/(4R\ST)))^2 }\right \}^{-1/2}.
\label{espitzer_a}
\end{equation}

\iffigscl
\begin{figure}
\includegraphics[width=\columnwidth]{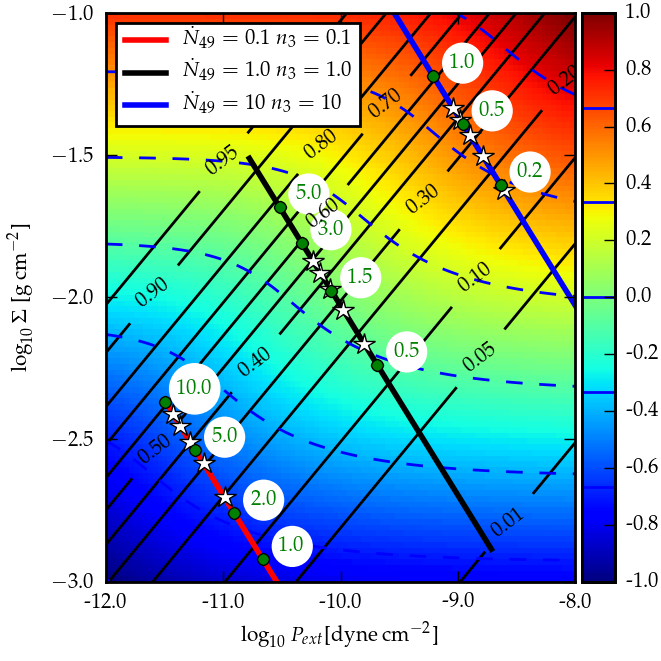}
\caption{
Evolutionary paths for shell walls around expanding \HII regions. 
The paths with ($\dot{N}_{49}$, $n_{3}$) = ($0.1$, $0.1$), 
($\dot{N}_{49}$, $n_{3}$) = ($1$, $1$) and ($\dot{N}_{49}$, $n_{3}$) = ($10$, $10$)
are shown by red, black and blue lines, respectively. 
Green circles with numbers give time in Myr since the expansion started. 
White asterisks represent the instants when the integral in \eq{efrg_integral} equals 0, 0.25, 0.50, 0.75 and 1.0. 
Parameter $A$ is shown by black labelled contours. 
Colourscale and dashed blue contours represent logarithm of the integrand in \eq{efrg_integral} expressed in Myr$^{-1}$.
The plot is constructed for $c_{s2} = 1$.
}
\label{fpsplane}
\end{figure}
\else
\fi

Evolution of $P\EXT$, $\Sigma$ and $A$ for walls of shells powered by sources with $\dot{N}_{49} = 0.1$ (approximately an O9V star \citep{ms05}), 
$\dot{N}_{49} = 1$ (O6V star) and $\dot{N}_{49} = 10$ (two O3V stars) expanding into a homogeneous 
ambient medium of particle density $n_{3} = 0.1$, $n_{3} = 1$ and $n_{3} = 10$, respectively, is shown in Figure \ref{fpsplane}. 
Plotted models are extremes in the sense that any evolutionary path for a shell within a set $\dot{N}_{49} \times n_{3}$ in range 
$(0.1, 10) \times (0.1, 10)$ lies in between them. 
The mentioned set $\dot{N}_{49} \times n_{3}$ covers typical values for galactic \HII regions. 
Time $t\subO$ as determined from \eq{etime_t0} below is plotted as the first asterisk on the evolutionary path. 
Following asterisks represent instants when integral in \eq{efrg_integral} reaches 0.25, 0.5, 0.75 and 1.0. 
They show that fragmentation proceeds in pressure dominated case.

Substituting $A$ from \eq{espitzer_a} for $t$ in \eq{efrg_integral} gives
\begin{equation}
\frac{0.81 c\subS}{\left \{(G  \rho\ACC)^3 R\ST^6 c_{II}^8 \right \}^{1/14}}  = \int_{A\subO}^{A_{1}} \frac{(1 - A^2)^{-25/14}}{n\EFOLD(A) A^{3/7}} \dd A,
\label{efrg_integral_a}
\end{equation}
where $A\subO = A(t\subO)$ and $A_1 = A(t_1)$ denote the value of parameter $A$ at the beginning and end of fragmentation, respectively.

\iflabs
\textit{Choice of $t_0$}
\else\fi
We adopt $t\subO$ as the earliest time when self--gravity of a fragment containing one 
Jeans mass (its radius is $R\JEANS = \sqrt{M\JEANS/(\pi \Sigma)}$) dominates stretching, i.e. $G\Sigma = R\JEANS (4/(7t\subO))^2$ (see appendix A in \citet{wb94a})
\footnote{Our choice of $t\subO$ is an approximation since for smaller fragments with $\lambda\DRMAX$, $\lambda\DRMAX < R\JEANS$ 
self--gravity dominates stretching earlier than at $t\subO$. However, the results (\eq{efit_shell}) are weakly dependent on $t\subO$; 
Any numerical coefficients in \eq{efit_shell} would differ by less than 20 \% if we used $t\subO = 0$ instead of \eq{etime_t0}.}
%
. We neglect the pressure gradient term since the unstable objects form via coalescence when the role of 
pressure support is already lost. Thus, 
\begin{equation}
t\subO = 0.76 \; \rmn{Myr} \; c_{s2}^{\frac{28}{37}} \; n_3^{-\frac{33}{74}} \; \dot{N}_{49}^{-\frac{4}{37}}.
\label{etime_t0}
\end{equation}

\subsection{The properties of fragments formed in a shell swept up by an expanding \HII region}

\iflabs
\textit{Numerical results}
\else\fi
Integrating \eq{efrg_integral_a} we find $A(t_1)$ and from \eq{espitzer_a} fragmenting time $t_1$. 
We evaluate integral \eq{efrg_integral_a} for $c\subS = 0.2 \rmn{km/s}$ on a set of $\dot{N}_{49} \times n_{3}$ 
in range $(0.1, 10) \times (0.1, 10)$, and by polynomial fitting we find
\begin{subequations}
\begin{align}
t\FRG &= 2.4 \; \Myr \; n_3^{-0.43} \; \dot{N}^{-0.12}_{49}  \label{eft} \\
M\FRG &= 3.1 \; \Msun \; n_3^{-0.41} \; \dot{N}^{-0.16}_{49} \label{efm}  \\
r\FRG &= 0.12 \; \Pc \; n_3^{-0.44} \; \dot{N}^{-0.11}_{49} \\
A\FRG &= 0.51 \; n_3^{0.04} \; \dot{N}^{-0.08}_{49} \label{efa} \\
\Sigma\FRG &= 63 \; \Sd \; n_3^{0.46} \; \dot{N}^{0.08}_{49} \label{efsd} \\
R\FRG &= 7.7  \; \rmn{pc} \; n_3^{-0.53} \; \dot{N}^{0.08}_{49},
\end{align}
\label{efit_shell}
\end{subequations}
where $M\FRG$ is fragment mass, $r\FRG$ fragment radius, and $A\FRG$, $R\FRG$ and $\Sigma\FRG$ are 
parameter $A$, shell radius and surface density at time $t\FRG$, respectively. 
Since we show in Sections \ref{snonlinear} and \ref{saccreting} that 
fragment mass of a pressure confined layer is comparable to the Jeans mass, 
we use the Jeans mass as the estimate of fragment mass in \eq{efm}. 
We compare \eq{efit_shell} with results of previous studies in Section \ref{sdsf}.


\iflabs
\textit{Fragmentation occurs in pressure dominated case}
\else\fi
Throughout this section, we have assumed that the shell fragments in external pressure dominated case.
The highest value attained by $A$ in \eq{efa} justifies this assumption ($A\FRG \la 0.6$) unless $n_3$ 
is either very high or $\dot{N}_{49}$ very low.

\iflabs
\textit{restriction to pressure dominated case}
\else\fi
We restrict our study to sound velocity $c_{s2} = 1$ (i.e. shell temperature $10 \, \mathrm{K}$) to fulfill the condition $A\FRG \la 0.6$ for realistic values of $\dot{N}_{49}$ and $n_3$ 
because the area in $(\dot{N}_{49}, n_3)$ plane where the condition is met decreases as $c\subS$ increases. 
Since fragmentation of self--gravity dominated shells is more complicated (time dependent $\rho_0$, 
gravitational focusing on emerging fragments, \dots), it is not clear what is the proxy for fragment mass in this case and we do not study it here.

\section{Discussion}

\subsection{Comparison of numerical with analytic dispersion relations}


\label{sdpagi}

\iflabs
\textit{Comparison with analytical models}
\else\fi
The simulations in linear regime (when perturbed surface density $\Sigma_1$ is smaller than unperturbed $\Sigma\subO$) clearly reproduce 
the semi--analytical estimate E78 in both self--gravity ($A = 0.99$) and pressure dominated ($A=0.18$) cases (Fig. \ref{fdisp_relat}). 
We do not detect any systematic difference from E78 confirming correctness of the derivation done by \citet{ee78}. 
Although W10 provides a good approximation to the range of unstable wavenumbers 
in the self--gravity dominated case, it underestimates the highest unstable wavenumber $k\MAX$ for $A \la 0.5$ (Fig. \ref{fdepona}). 
V83 overestimates $k\MAX$ by a factor $2$ for $A=1$, predicts correct $k\MAX$ for $A \simeq 0.6$ and underestimates $k\MAX$ for lower $A$. 

\iflabs
\textit{Comparison with prev. work}
\else\fi
Comparing to work of \citet{dw09} and \citet{wd10}, our non--accreting models are more suitable for testing dispersion relations of a layer 
since they have time independent $\Sigma\subO$ and no stretching. They are also completely free of Vishniac and Rayleigh--Taylor instabilities.
The lowest value of $A$ that \citet{wd10} reach is $0.42$ (bottom right panel in their fig. 7). 
Although their results favour W10 to V83, they can hardly distinguish W10 from E78.
We recover a less noisy dispersion relation and also reach lower value of $A$ where we can detect the differences between W10 and E78.

\iflabs
\textit{Simple model for perturbation}
\else\fi
Since W10 was proposed to include the influence of the external pressure, it may be surprising why W10 differs from 
our results (and therefore E78) when $A \la 0.5$. 
We illustrate the limitation of W10 on a simple model where the layer is perturbed by an overdensity 
in a form of a circular patch of radius $R_{_C}$ and central surface density $\Sigma_1$. The perturbed surface density decreases to zero from 
the centre to the edge at $R_{_C}$. 
We estimate its stability against lateral collapse by comparing self--gravity with pressure gradient between the
centre and the edge of the perturbation taken in the midplane. Then the condition for a lateral collapse is
\begin{equation}
G \beta \pi \Sigma_1 \ga \frac{P_1 - P\subO}{\rho\subO R_{_C}},
\label{epgeq_coll}
\end{equation}
where $P_1$ and $P\subO$ is the pressure in the centre and the edge of the perturbation, respectively.
Factor $\beta$ takes into account that the perturbation generates non--spherically symmetric gravitational field.
Factor $\beta$ is of order unity.
Substituting $\Sigma_1 + \Sigma\subO$ and $\Sigma\subO$ into \eq{erho0} and using \eq{ehalfth}, \eq{epgeq_coll} becomes 
$R_{_C} \beta \ga 2H\HALF$. Assuming $R_{_C} = \lambda/2$, \eq{epgeq_coll} takes the form 
\begin{equation}
\beta \pi/2 \ga k H\HALF.
\label{epgb}
\end{equation}
%
Therefore the model explains why the range of unstable wavelengths is of the order of the layer thickness for any $A$ as proposed by E78 and found in our simulations.

\iflabs
\textit{Reason why W10 fails}
\else\fi
The limit of W10 (\eq{pagidr}) in $A \to 0$ predicts a significantly narrower range of unstable wavelengths, $k H\HALF \leq 9 \pi^2 A^2 /20$. 
The difference is due to how a perturbation is treated in the models. 
Imposing a perturbation with positive surface density $\Sigma_1$ when $A$ is small, 
the layer thickens locally, so the majority of perturbed mass is deposited above the surface with very 
little density enhancement in the midplane. 
The layer is almost of constant density where variations in surface density are dominated by corrugations of the surface.
The weak dependence of $P_1$ on $\Sigma_1$ when $2P\EXT/(\pi G \Sigma\subO^2) \gg 1$ is apparent from \eq{erho0}. 
The density (pressure) gradient is much smaller than it would be 
if the perturbation were placed in the midplane, so the pressure gradient is easily overcame by self--gravity and substantially shorter wavelengths are unstable. 
\itiitext{In contrast}, this mechanism is absent in W10 where the perturbed mass is in the form of a homogeneous spheroid with no possibility to 
place the excess mass outside its surfaces. 
This assumption in W10 results in much higher difference between pressures $P\subO$ and $P_1$, 
which needs a more massive fragment (with longer $\lambda\DRMAX$) to be overcame with its self--gravity. 

In Section \ref{scol_time}, assuming that shells swept up by \HII regions are at temperature $10 {\rm K}$, we conclude 
that the shells fragment in pressure dominated case. 
In this case, the significant difference between E78 and the other two dispersion relations leads to a substantially different prediction for fragment masses. 
On the other hand, V83 or W10 may still provide rough estimates for fragment masses of large \HI shells, which are self--gravity dominated. 
Also note that E78 cannot be applied to layers forming at the interface between two colliding streams because such layers lack a surface with a contact discontinuity. 





\subsection{The effect of pre-existing density structure in the accretion flow}

\iflabs
\textit{Possible role of preexisting perturbations}
\else\fi
We assume that the perturbing amplitudes for both polychromatic and accreting simulations do not depend on $k$ (see Section \ref{spolyinit} for details), 
and that the accreting simulations accrete homogeneous medium. 
These assumptions significantly constrain the possible spectrum of perturbations.
This simple model was adopted in order to reduce the number of free parameters.
The real interstellar medium is highly inhomogeneous and when accreted, the 
resulting surface density enhancements have presumably different spectrum of perturbations. 
Since our fit giving the dependence of 
the number of e--folding times $n\EFOLD$ on $A$ is done for the spectrum we adopt, 
the form of $n\EFOLD = n\EFOLD(A)$ is presumably a function of the spectrum shape. 
In other words, the spectrum introduces another dimensionless parameters into \eq{edimless_time}.
The parameters may further alter the properties of the fragmenting shell (\eq{efit_shell}).

\iflabs
\textit{Merging with perturbations}
\else\fi
Accretion of even a single cloud could modify properties of final fragments substantially. 
For illustration, consider a spherically symmetric cloud.
When accreted on a homogeneous layer, the ensuing structure would be a superposition of the layer and a circular overdensity. 
We further assume that the circular cloud contains many Jeans masses.
This configuration is presumably prone to the edge effect \citep{bh04, pt12}, so it would tend to collapse laterally on its own 
timescale $t_{GC}$ proportional to its radius $r_{AC}$ and overdensity $\Sigma_1$ as $t_{GC} \simeq \sqrt{r_{AC}/(\pi G \Sigma_1)}$. 
Given that the layer contains preexisting perturbations of small amplitudes, the globally collapsing circular cloud would 
also fragment as a part of a layer and would break into pieces as we saw in Section \ref{snonlinear}. 

We suggest that external pressure could influence the result of fragmentation also in these circumstances if 
the circular cloud breaks into pieces before it collapsed globally. 
If the cloud is dominated by self--gravity, the fragments undergo an imminent collapse, so their radii shrink significantly 
diminishing probability of further coalescence. 
Consequently, the fragments would form a cluster in the center. 
If the cloud is dominated by external pressure, the fragments are still stable against gravitational collapse, so their radii 
are comparable to their distances, and they can coalesce. 
However, the coalescence rate would be significantly enhanced in comparison to the scenario we saw in Section \ref{snonlinear}, 
since the fragments would be attracted to the centre or swept up by the passing edge of the cloud. 
This probably leads to objects substantially more massive than the Jeans mass.
This issue may be an objective for future work.

\subsection{Fragmenting shells in other works}


\label{sdsf}

\iflabs
\textit{Possible advantages and drawbacks of our models}
\else\fi
When estimating properties of fragments forming in the shell, we use significantly simplified models.
The most important simplifications are the following: no deceleration, no ionising radiation penetrating into the shell, no complicated thermodynamics, 
and homogeneous medium into which the shell expands. 
Vertical acceleration of the unperturbed layer has reflective symmetry relative to the midplane; this assumption is violated 
in decelerating shells where it changes the dispersion relation as is shown by \citet{ii11b}.
The role of stretching is probably not crucial as estimated by \citet{wb94b}.
These phenomena may influence the course of fragmentation significantly and possibly modify or even suppress coalescence driven collapse in real shells. 

Nevertheless, our work improves the model proposed by \citet{wb94b}, which adopts the same simplified assumptions, 
since we use more appropriate estimates for mass of fragments and relevant timescales. 
Taking into account approximations we made, we emphasize that \eq{efit_shell} is rather 
an application of our results for layers than the definite answer to the problem of propagating star formation due to photoionising feedback.
Assuming temperature of $10 \mathrm{K}$ in the shells, \eq{efit_shell} predicts 
that shell fragmentation tends to form objects of too low mass for star formation to propagate.

\iflabs
\textit{Comparison with previous works}
\else\fi
The dependence of our estimate \eq{efit_shell} on $\dot{N}_{49}$ and $n_3$ 
is very close to the results provided by \citet{wb94b} and \citet{ii11a} (with sound velocity $c_{s2} = 1$).
Compared to \citet{wb94b}, \eq{eft} predicts a slightly later fragmenting time ensuing in a larger shell radius 
(numerical constants they provide are $1.56  \Myr$ and $5.8  \Pc$). 
However, our model predicts much smaller fragment masses ($3.1  \Msun$ vs. $23  \Msun$) and smaller fragment radii ($0.12  \Pc$ vs. $0.41  \Pc$).
This result is the consequence of taking as a proxy for fragment mass the Jeans mass for a significantly compressed layer instead of
the Jeans mass for not compressed layer as done by \citet{wb94b}.

\iflabs
\textit{Comparison with D07}
\else\fi
In order to test the numerical constants in formulae derived by \citet{wb94b}, \citet{db07} select one point in $\dot{N}_{49}, n_3$ plane, 
$(\dot{N}_{49}, n_3) = (1.0, 0.2)$ and follow its evolution by an SPH simulation.
They find an approximate agreement with the work of \citet{wb94b}.
However, comparing density of their shell $\sim 10^{-21} \mathrm{g \; cm}^{-3}$ with the analytic value which is approx. $50 \times$ higher, 
we suspect that their resolution is not enough to resolve the shell vertically. 
The low--density in the shell leads to fragments of artificially high mass.

\iflabs
\textit{Comparison with II11}
\else\fi
On the other hand, our mass estimate in \eq{efm} is close to the value $3.5 \Msun$ found by \citet{ii11a} based on monochromatic SPH simulations.
We cannot compare fragmenting time since they use a different definition of the quantity.


\section{Conclusions}

We perform 3D hydrodynamic simulations of self--gravitating isothermal layers in order to investigate their dispersion relations 
and subsequent fragmentation. We apply the results of fragmenting and accreting layers to estimate typical masses and fragmenting time for 
a shell swept up around an \HII region.

We find that if the perturbations are small, thermal pressure confined layers fragment in an excellent agreement with the dispersion relation 
proposed by \citet{ee78} both in the self--gravity and pressure dominated case. 
PAGI \citep{wd10} is a good approximation and thin shell \citep{v83} a rough approximation in the self--gravity dominated case. 
However, the latter two are inappropriate for the pressure dominated case ($A \la 0.5$; \eq{a}) 
because they do not take into account surface corrugations.
Since the shells around \HII regions 
presumably fragment in the pressure dominated case, we suggest that both PAGI and thin shell dispersion relations 
are not appropriate for their description.

Fragmentation in the non--linear regime proceeds in two qualitatively different ways depending on parameter $A$. 
When the layer is dominated by self--gravity, fragments form by a monolithic collapse. 
When the layer is dominated by external pressure, it fragments in two steps; firstly it breaks into 
small gravitationally stable objects. It is an interesting feature because purely 
gravitational instability in planar geometry can form objects which are themselves stable against gravitational collapse. 
Then, the fragments continuously merge until they form an unstable 
object which eventually collapses (coalescence driven collapse). 
Neither of the dispersion relations presented can predict correctly properties of gravitationally bound fragments for a pressure dominated layer.
However, a rough estimate for their mass is the Jeans mass in the midplane of the layer.

In the non--linear regime, we investigate the possibility that the layer self--organises and forms a regular pattern on its surface. 
We seek our standard models covering various degree of external pressure confinement ($A = 0.18$, $A=0.60$, $A = 0.99$), and find no evidence for this scenario. 
In addition, we obtain the same result with two models designed to enhance pattern formation.
We note that our method is able to find any regular pattern, not only a hexagonal pattern as is suggested to form by analytic work of \citet{f96}.
Instead of regular patterns, we observe formation of randomly oriented filamentary--like structures in agreement with \citet{mn87a, mn87b}.

For pressure dominated layers, we substitute boundary conditions on one surface from thermal pressure for an accreting homogeneous medium. 
The dispersion relation for accreting layers has the range of unstable wavenumbers extended towards higher values of $k$ in comparison to the 
dispersion relation for thermal pressure confined layers. 
The highest growth rate is by factor $\simeq 2$ higher. 
In the non--linear regime of fragmentation, accreting layers also undergo monolithic or coalescence driven collapse depending 
on the importance of confining pressure during the major part of fragmentation. 
Layers with high accreting rate become earlier dominated by self--gravity and collapse monolithically, while 
layers with lower accreting rate remain dominated by external pressure and fragment via coalescence driven collapse. 
Fragmentation occurs at time comparable to thermal pressure confined layers with the same instantaneous surface density and ambient pressure.
The mass of gravitationally bound fragments is again comparable to the Jeans mass.

We use our results from accreting layers to estimate fragment properties of a shell swept--up around an expanding \HII region. 
For typical density ($10^3 \Cmiii$) and temperature ($10 \; \rmn{K}$) in molecular clouds, the fragmentation 
is accomplished while the shell is still dominated by the external pressure. 
This leads to fragment masses $\simeq 3 \Msun$.
Stars formed of fragments of this mass are not able to ignite new \HII regions.  
This indicates that for star formation to propagate, 
either higher temperature in shells or a different scenario (e.g. geometry) is required.

\section*{Acknowledgments}

The authors like to thank the referee, Sven Van Loo for constructive comments which helped to significantly improve the paper. 
FD, RW and JP acknowledge support from the Czech Science Foundation  grant  P209/12/1795 and by the project RVO:67985815.
The presented simulations were performed on IT4I facilities supported by The Ministry of Education, Youth and
Sports from the Large Infrastructures for Research, Experimental Development and Innovations
project „IT4Innovations National Supercomputing Center – LM2015070“.

\bibliographystyle{mn2e}
\bibliography{gi_layer}

\bsp

\appendix
\section{Clump finding algorithm}

\iffigscl
\begin{figure}
\includegraphics[width=\columnwidth]{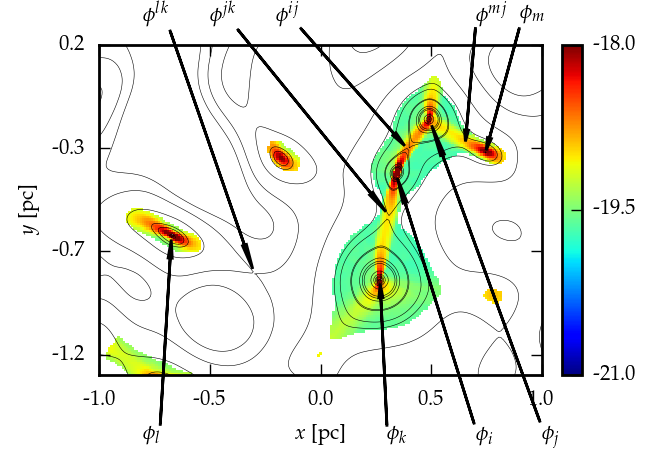}
\caption{Position of saddles and potential minima to illustrate our clump finding algorithm. 
Potential minima and saddle points are marked by subscripts and superscripts, respectively. 
Equipotential curves are plotted by thin solid lines, cells identified as bound are plotted by colour corresponding to their density. 
Note that only selected minima and saddles are marked.}
\label{fclump_find}
\end{figure}
\else
\fi

\label{APP:CLUMPFIND}

In order to determine the fragmenting time and mass of emerging bound objects, we need to identify gravitationally bound entities. 
We search for a gravitationally bound object around each minimum in gravitational potential in the computational domain. 
The gravitational potential is a sum of the gaseous potential as determined by {\sc FLASH} code and the potential 
due to sink particles.
When investigating a particular minimum $i$ at potential $\phi_{i}$ (Figure \ref{fclump_find}), 
we draw a sphere of the sink particle radius $r\SINK$ with centre at the potential minimum, 
and check whether the sphere contains at least one grid cell with negative binding energy $e\CELL$.
The binding energy of a cell is
\begin{eqnarray}
e\CELL & = & \frac{1}{2} m\CELL (v\CELL - \VCL)^2 + \frac{3 m\CELL cs\CELL^2}{2} + \nonumber \\
&&\frac{1}{2} m\CELL (\phi\CELL - \phi^{ij}), 
\label{ecell}
\end{eqnarray}
where $m\CELL$, $\phi\CELL$, $v\CELL$ and $cs\CELL$ are mass, potential, bulk velocity and sound velocity of the cell. 
The kinetic energy is corrected by velocity of the mass centre $\VCL$ of the bound object, which is 
composed of all bound grid cells.

The gravitational binding energy is related to the potential at the saddle point $\phi^{ij}$ 
(the second letter in the superscript denotes the deeper minimum, i.e. $\phi_{j} < \phi_{i}$). 
The saddle for minimum $i$ is a pair of two neighbouring cells
at the lowest potential $\phi\CELL$ where a path in potential from one cell monotonically descends 
to minimum $i$ and a path from the other cell monotonically descends to another, deeper 
minimum $j$ (see Fig. \ref{fclump_find}). 
Mixed boundary conditions are taken into account, so the path can lead through a periodic boundary. 
No cell with $\phi\CELL > \phi^{ij}$ can be bound to the minimum $i$.

If at least one bound cell is found, the radius of the sphere is 
increased by one grid cell size and condition \eq{ecell} is applied to any cell in the spherical shell 
between the new and previous searching radius. 
The procedure is repeated until the sphere 
exceeds the saddle point, or until it exceeds the cloud boundary, i.e. all cells in the spherical shell are unbound. 
If sink particles are present, they are very close to local potential minima and their mass is added to the mass budget of the 
bound object identified around the minimum.

The computational domain typically contains many objects. 
Considerable number of the objects are interacting and overlapping. 
The algorithm firstly sorts the minima in ordering of decreasing potential 
and starts searching around the minimum with the highest value of potential and continues in direction of decreasing potential. 
This searching direction assures that the objects with minima at higher potential levels, which 
are usually situated close to more massive objects with minima at lower potential levels, are investigated 
before so that their mass is properly assigned to the lower mass objects. 
Any cell identified as bound is marked, and can not be accessed later when investigating a deeper minimum.

We illustrate some of the features of the algorithm on Fig. \ref{fclump_find}. 
Note that the ordering of potential minima ($\phi_m > \phi_l > \phi_i > \phi_j > \phi_k$) is not reflected by ordering of saddles 
$\phi^{lk} > \phi^{jk} > \phi^{mj} > \phi^{ij}$. 
Only cells around $i$ with $\phi < \phi^{ij}$ can be assigned to minimum $i$. 
Since the saddle $\phi^{jk}$ for object $j$ is at a higher equipotential surface then previous saddle $\phi^{ij}$, many cells 
situated close to $i$, which were not previously assigned to minimum $i$ are now identified as gravitationally bound and assigned to $j$. 
Cells situated around a shallow minimum at $m$, where no bound object was found previously, are checked if they are bound to $j$. 
The algorithm is able to find objects both filling and underfilling their critical equipotentials (e.g. $i$ and $l$) and 
also identify highly non--spherically symmetric objects (e.g. $j$).

%

\label{lastpage}

\end{document}